%% file: paper324_final.tex
\documentstyle[10pt,epsf,epsfig,hangcaption,xspace,amssymb,amsfonts,
dp_delphititle,lineno]{dp_delphi}

\makeindex
\def\DpPaperGroup{EP}
\def\DpPaperRef{2003-093}
\def\DpDate{18 December 2003}
\def\DpAuthors{DELPHI Collaboration}
\def\DpSubmit{(Accepted by Euro. Phys. J. C)}
\def\DpTitle{{ Photon Events with Missing Energy in e$^+$e$^-$ Collisions 
at \boldmath $\sqrt{s}~=$~130 to 209~GeV}}
\def\DpComment{   } 
\def\DpEMail{  }

\newcommand {\ee} {\rm{e}^+\rm{e}^-}

\newcommand {\eenng} {e^+e^-\rightarrow \nu\bar{\nu}\gamma}
\newcommand {\eenngg} {e^+e^-\rightarrow \nu\bar{\nu}\gamma\gamma(\gamma)}
\newcommand {\eegg} {e^+e^-\rightarrow \gamma\gamma}
\newcommand {\eeGG} {e^+e^-\rightarrow\tilde{G}\tilde{G}\gamma}
\newcommand {\eeGGg} {e^+e^-\rightarrow\tilde{G}\tilde{\chi}^0_1\rightarrow\tilde{G}\tilde{G}\gamma}
\newcommand {\eeGgGg} {e^+e^-\rightarrow\tilde{\chi}^0_1\tilde{\chi}^0_1\rightarrow\tilde{G}\gamma\tilde{G}\gamma} 
\newcommand {\eeXXg} {e^+e^-\rightarrow\tilde{\chi}^0_1\tilde{\chi}^0_2\rightarrow\tilde{\chi}^0_1\tilde{\chi}^0_1\gamma} 
\newcommand {\eeXgXg} {e^+e^-\rightarrow\tilde{\chi}^0_2\tilde{\chi}^0_2\rightarrow\tilde{\chi}^0_1\gamma\tilde{\chi}^0_1\gamma}

\makeatletter
\renewcommand{\theenumi}{\Roman{enumi}}

\renewcommand{\theenumii}{\Alph{enumii}}

\renewcommand{\p@enumii}{\theenumi--}
\makeatother

\begin{document}
\makeatletter
\makeatother

\begin{titlepage}
\pagenumbering{roman}

\CERNpreprint{\DpPaperGroup}{\DpPaperRef} 
\date{{\small\DpDate}} 
\title{\DpTitle} 
\address{\DpAuthors} 

\begin{shortabs} 
\noindent
%
\noindent

The production of single- and multi-photon events has been studied
in the reaction $e^+e^-\rightarrow\gamma(\gamma)+invisible~particles$.
The data collected with the DELPHI detector during the years 1999 and 2000 
at centre-of-mass energies between 191~GeV and 209~GeV was combined with
earlier data to search for phenomena beyond the Standard Model. 
The measured number of light neutrino families was consistent with three and the 
absence of an excess of events beyond that predicted by the Standard Model
processes was used to set limits on new physics. Both model-independent searches
and searches for new processes predicted by supersymmetric and extra-dimensional
models have been made. Limits on new non-standard model interactions between 
neutrinos and electrons were also determined.
\end{shortabs}

\vfill

\begin{center}
\DpSubmit \ \\ 
\DpComment \ \\
\DpEMail \ \\
\end{center}

\vfill
\clearpage

\headsep 10.0pt

\addtolength{\textheight}{10mm}
\addtolength{\footskip}{-5mm}
\begingroup
%
\newcommand{\DpName}[2]{\hbox{#1$^{\ref{#2}}$},\hfill}
\newcommand{\DpNameTwo}[3]{\hbox{#1$^{\ref{#2},\ref{#3}}$},\hfill}
\newcommand{\DpNameThree}[4]{\hbox{#1$^{\ref{#2},\ref{#3},\ref{#4}}$},\hfill}
\newskip\Bigfill \Bigfill = 0pt plus 1000fill
\newcommand{\DpNameLast}[2]{\hbox{#1$^{\ref{#2}}$}\hspace{\Bigfill}}

%
\footnotesize
\noindent
\DpName{J.Abdallah}{LPNHE}
\DpName{P.Abreu}{LIP}
\DpName{W.Adam}{VIENNA}
\DpName{P.Adzic}{DEMOKRITOS}
\DpName{T.Albrecht}{KARLSRUHE}
\DpName{T.Alderweireld}{AIM}
\DpName{R.Alemany-Fernandez}{CERN}
\DpName{T.Allmendinger}{KARLSRUHE}
\DpName{P.P.Allport}{LIVERPOOL}
\DpName{U.Amaldi}{MILANO2}
% ...
\normalsize
\endgroup

\titlefoot{Department of Physics and Astronomy, Iowa State
     University, Ames IA 50011-3160, USA
    \label{AMES}}
\titlefoot{Physics Department, Universiteit Antwerpen,
     Universiteitsplein 1, B-2610 Antwerpen, Belgium \\
     \indent~~and IIHE, ULB-VUB,
     Pleinlaan 2, B-1050 Brussels, Belgium \\
     \indent~~and Facult\'e des Sciences,
     Univ. de l'Etat Mons, Av. Maistriau 19, B-7000 Mons, Belgium
    \label{AIM}}
\titlefoot{Physics Laboratory, University of Athens, Solonos Str.
     104, GR-10680 Athens, Greece
    \label{ATHENS}}
\titlefoot{Department of Physics, University of Bergen,
     All\'egaten 55, NO-5007 Bergen, Norway
    \label{BERGEN}}
\titlefoot{Dipartimento di Fisica, Universit\`a di Bologna and INFN,
     Via Irnerio 46, IT-40126 Bologna, Italy
    \label{BOLOGNA}}
\titlefoot{Centro Brasileiro de Pesquisas F\'{\i}sicas, rua Xavier Sigaud 150,
     BR-22290 Rio de Janeiro, Brazil \\
     \indent~~and Depto. de F\'{\i}sica, Pont. Univ. Cat\'olica,
     C.P. 38071 BR-22453 Rio de Janeiro, Brazil \\
     \indent~~and Inst. de F\'{\i}sica, Univ. Estadual do Rio de Janeiro,
     rua S\~{a}o Francisco Xavier 524, Rio de Janeiro, Brazil
    \label{BRASIL}}
\titlefoot{Coll\`ege de France, Lab. de Physique Corpusculaire, IN2P3-CNRS,
     FR-75231 Paris Cedex 05, France
    \label{CDF}}
\titlefoot{CERN, CH-1211 Geneva 23, Switzerland
    \label{CERN}}
\titlefoot{Institut de Recherches Subatomiques, IN2P3 - CNRS/ULP - BP20,
     FR-67037 Strasbourg Cedex, France
    \label{CRN}}
\titlefoot{Now at DESY-Zeuthen, Platanenallee 6, D-15735 Zeuthen, Germany
    \label{DESY}}
\titlefoot{Institute of Nuclear Physics, N.C.S.R. Demokritos,
     P.O. Box 60228, GR-15310 Athens, Greece
    \label{DEMOKRITOS}}
\titlefoot{FZU, Inst. of Phys. of the C.A.S. High Energy Physics Division,
     Na Slovance 2, CZ-180 40, Praha 8, Czech Republic
    \label{FZU}}
\titlefoot{Dipartimento di Fisica, Universit\`a di Genova and INFN,
     Via Dodecaneso 33, IT-16146 Genova, Italy
    \label{GENOVA}}
\titlefoot{Institut des Sciences Nucl\'eaires, IN2P3-CNRS, Universit\'e
     de Grenoble 1, FR-38026 Grenoble Cedex, France
    \label{GRENOBLE}}
\titlefoot{Helsinki Institute of Physics, P.O. Box 64,
     FIN-00014 University of Helsinki, Finland
    \label{HELSINKI}}
\titlefoot{Joint Institute for Nuclear Research, Dubna, Head Post
     Office, P.O. Box 79, RU-101 000 Moscow, Russian Federation
    \label{JINR}}
\titlefoot{Institut f\"ur Experimentelle Kernphysik,
     Universit\"at Karlsruhe, Postfach 6980, DE-76128 Karlsruhe,
     Germany
    \label{KARLSRUHE}}
\titlefoot{Institute of Nuclear Physics PAN,Ul. Radzikowskiego 152,
     PL-31142 Krakow, Poland
    \label{KRAKOW1}}
\titlefoot{Faculty of Physics and Nuclear Techniques, University of Mining
     and Metallurgy, PL-30055 Krakow, Poland
    \label{KRAKOW2}}
\titlefoot{Universit\'e de Paris-Sud, Lab. de l'Acc\'el\'erateur
     Lin\'eaire, IN2P3-CNRS, B\^{a}t. 200, FR-91405 Orsay Cedex, France
    \label{LAL}}
\titlefoot{School of Physics and Chemistry, University of Lancaster,
     Lancaster LA1 4YB, UK
    \label{LANCASTER}}
\titlefoot{LIP, IST, FCUL - Av. Elias Garcia, 14-$1^{o}$,
     PT-1000 Lisboa Codex, Portugal
    \label{LIP}}
\titlefoot{Department of Physics, University of Liverpool, P.O.
     Box 147, Liverpool L69 3BX, UK
    \label{LIVERPOOL}}
\titlefoot{Dept. of Physics and Astronomy, Kelvin Building,
     University of Glasgow, Glasgow G12 8QQ
    \label{GLASGOW}}
\titlefoot{LPNHE, IN2P3-CNRS, Univ.~Paris VI et VII, Tour 33 (RdC),
     4 place Jussieu, FR-75252 Paris Cedex 05, France
    \label{LPNHE}}
\titlefoot{Department of Physics, University of Lund,
     S\"olvegatan 14, SE-223 63 Lund, Sweden
    \label{LUND}}
\titlefoot{Universit\'e Claude Bernard de Lyon, IPNL, IN2P3-CNRS,
     FR-69622 Villeurbanne Cedex, France
    \label{LYON}}
\titlefoot{Dipartimento di Fisica, Universit\`a di Milano and INFN-MILANO,
     Via Celoria 16, IT-20133 Milan, Italy
    \label{MILANO}}
\titlefoot{Dipartimento di Fisica, Univ. di Milano-Bicocca and
     INFN-MILANO, Piazza della Scienza 2, IT-20126 Milan, Italy
    \label{MILANO2}}
\titlefoot{IPNP of MFF, Charles Univ., Areal MFF,
     V Holesovickach 2, CZ-180 00, Praha 8, Czech Republic
    \label{NC}}
\titlefoot{NIKHEF, Postbus 41882, NL-1009 DB
     Amsterdam, The Netherlands
    \label{NIKHEF}}
\titlefoot{National Technical University, Physics Department,
     Zografou Campus, GR-15773 Athens, Greece
    \label{NTU-ATHENS}}
\titlefoot{Physics Department, University of Oslo, Blindern,
     NO-0316 Oslo, Norway
    \label{OSLO}}
\titlefoot{Dpto. Fisica, Univ. Oviedo, Avda. Calvo Sotelo
     s/n, ES-33007 Oviedo, Spain
    \label{OVIEDO}}
\titlefoot{Department of Physics, University of Oxford,
     Keble Road, Oxford OX1 3RH, UK
    \label{OXFORD}}
\titlefoot{Dipartimento di Fisica, Universit\`a di Padova and
     INFN, Via Marzolo 8, IT-35131 Padua, Italy
    \label{PADOVA}}
\titlefoot{Rutherford Appleton Laboratory, Chilton, Didcot
     OX11 OQX, UK
    \label{RAL}}
\titlefoot{Dipartimento di Fisica, Universit\`a di Roma II and
     INFN, Tor Vergata, IT-00173 Rome, Italy
    \label{ROMA2}}
\titlefoot{Dipartimento di Fisica, Universit\`a di Roma III and
     INFN, Via della Vasca Navale 84, IT-00146 Rome, Italy
    \label{ROMA3}}
\titlefoot{DAPNIA/Service de Physique des Particules,
     CEA-Saclay, FR-91191 Gif-sur-Yvette Cedex, France
    \label{SACLAY}}
\titlefoot{Instituto de Fisica de Cantabria (CSIC-UC), Avda.
     los Castros s/n, ES-39006 Santander, Spain
    \label{SANTANDER}}
\titlefoot{Inst. for High Energy Physics, Serpukov
     P.O. Box 35, Protvino, (Moscow Region), Russian Federation
    \label{SERPUKHOV}}
\titlefoot{J. Stefan Institute, Jamova 39, SI-1000 Ljubljana, Slovenia
     and Laboratory for Astroparticle Physics,\\
     \indent~~Nova Gorica Polytechnic, Kostanjeviska 16a, SI-5000 Nova Gorica, Slovenia, \\
     \indent~~and Department of Physics, University of Ljubljana,
     SI-1000 Ljubljana, Slovenia
    \label{SLOVENIJA}}
\titlefoot{Fysikum, Stockholm University,
     Box 6730, SE-113 85 Stockholm, Sweden
    \label{STOCKHOLM}}
\titlefoot{Dipartimento di Fisica Sperimentale, Universit\`a di
     Torino and INFN, Via P. Giuria 1, IT-10125 Turin, Italy
    \label{TORINO}}
\titlefoot{INFN,Sezione di Torino, and Dipartimento di Fisica Teorica,
     Universit\`a di Torino, Via P. Giuria 1,\\
     \indent~~IT-10125 Turin, Italy
    \label{TORINOTH}}
\titlefoot{Dipartimento di Fisica, Universit\`a di Trieste and
     INFN, Via A. Valerio 2, IT-34127 Trieste, Italy \\
     \indent~~and Istituto di Fisica, Universit\`a di Udine,
     IT-33100 Udine, Italy
    \label{TU}}
\titlefoot{Univ. Federal do Rio de Janeiro, C.P. 68528
     Cidade Univ., Ilha do Fund\~ao
     BR-21945-970 Rio de Janeiro, Brazil
    \label{UFRJ}}
\titlefoot{Department of Radiation Sciences, University of
     Uppsala, P.O. Box 535, SE-751 21 Uppsala, Sweden
    \label{UPPSALA}}
\titlefoot{IFIC, Valencia-CSIC, and D.F.A.M.N., U. de Valencia,
     Avda. Dr. Moliner 50, ES-46100 Burjassot (Valencia), Spain
    \label{VALENCIA}}
\titlefoot{Institut f\"ur Hochenergiephysik, \"Osterr. Akad.
     d. Wissensch., Nikolsdorfergasse 18, AT-1050 Vienna, Austria
    \label{VIENNA}}
\titlefoot{Inst. Nuclear Studies and University of Warsaw, Ul.
     Hoza 69, PL-00681 Warsaw, Poland
    \label{WARSZAWA}}
\titlefoot{Fachbereich Physik, University of Wuppertal, Postfach
     100 127, DE-42097 Wuppertal, Germany
    \label{WUPPERTAL}}
\addtolength{\textheight}{-10mm}
\addtolength{\footskip}{5mm}
\clearpage

\headsep 30.0pt
\end{titlepage}

%
\pagenumbering{arabic} 
\setcounter{footnote}{0} %
\large
\section{Introduction}

The DELPHI experiment has previously reported studies of events at
centre-of-mass energies up to 189.2 GeV in which only one photon or two
acoplanar\footnotemark ~photons were 
produced~\cite{delphi_lep1,delphi_130,delphi_chargino,delphi_gg,delphi_189}.  
The other LEP experiments have reported similar
studies~\cite{LEP_results}.  The present paper combines all DELPHI
single-photon data recorded at 180-209 GeV and all acoplanar two-photon data 
recorded at 130-209 GeV to obtain the most stringent limits
on new physics.  The paper improves and supersedes the earlier studies
made at lower energies.

\footnotetext{The acoplanarity is here defined as
$180^{\circ}-\Phi^{T}_{12}$, 
where $\Phi^T_{12}$ is the angle between the two photons in the plane
perpendicular to the beam axis.}

At LEP2, the Standard Model predicts that events with one or more
photons and invisible particles are produced mainly by the
reaction $e^+e^-\rightarrow\nu\bar{\nu}\gamma(\gamma)$ which receives
a contribution from Z-exchange in the $s$-channel with single- or
multi-photon emission from the initial state electrons and from the
$t$-channel $W$-exchange, with the photon(s) radiated from the beam
electrons or the exchanged $W$.  Beyond the Standard Model,
contributions to the $\gamma~+~missing~energy$ final state could come
from a new generation of neutrinos or from the production of
some new particle, stable or unstable, weakly interacting or decaying
into a photon and an invisible particle.

A measurement of the cross-section of the process $e^+e^- \rightarrow
\nu \bar{\nu}\gamma$ determines the number of light neutrino
generations, $N_{\nu}$.  DELPHI has previously reported measurements
of $N_{\nu}$ using both LEP1~\cite{delphi_lep1} and LEP2~\cite{delphi_189} 
single-photon data and this
measurement has now been re-evaluated with all 180-209~GeV DELPHI data.

The $e^+e^- \rightarrow \nu \bar{\nu}\gamma$ cross-section can also be
used to calculate limits on new neutrino interactions with electrons,
beyond what is predicted by the Standard Model. A set of limits
on the parameter space for such new neutral-current interactions
is presented in this paper.

A search for new particles can be done both
in the framework of specific models and in a model-independent way.  
The event topology with one or two photons and missing energy
can be used to look for the production of either new invisible
particles tagged by initial-state radiation (ISR) or the production of
invisible particles in association with a photon.  The 
model-independent single-photon searches described in this paper take both of these
possibilities into account.  The first search is for the production of
a new hypothetical particle tagged by an ISR-photon,
$e^+e^-\rightarrow X\gamma$. In this analysis, the new particle, $X$, was
assumed to be either stable and weakly interacting or to be
decaying into invisible decay products.  The second model-independent
search consisted of looking for and setting cross-section limits on
the process $e^+e^-\rightarrow XY \rightarrow YY\gamma$ where $X$ is a
hypothetical new neutral particle which decays radiatively to $Y$
which is a stable weakly interacting new neutral particle.

A specific case in which photons are produced in association
with a new particle was addressed in the search for gravitons via the process
$e^+e^-\rightarrow\gamma G$.  The existence of this process has been
predicted by some string models~\cite{grav1,grav2} in which gravity is
allowed to propagate in a space with more dimensions than the usual
four space-time dimensions.

Various theories of supersymmetry (SUSY) predict the existence of new
particles which would produce final states with missing energy and one
or several photons.  Different assumptions about the SUSY breaking
mechanism lead to two main search scenarios in which the Lightest
Supersymmetric Particle (LSP) is either the gravitino ($\tilde{G}$) or
the neutralino ($\tilde{\chi}^0_1$).

Gauge-Mediated Supersymmetry Breaking (GMSB) models~\cite{GMSB}
typically predict that the gravitino is the LSP with a mass less than
a few hundred eV/c$^2$ and that the neutralino or the slepton is the
Next-to-Lightest Supersymmetric Particle (NLSP).  If the neutralino is
the NLSP it decays into $\tilde{G}\gamma$ with the gravitino being
essentially massless and undetectable and it is possible to search for
both single-photon and multi-photon production via the processes
$\eeGGg$ and $\eeGgGg$.  The cross-section for the single-photon
process is proportional to $1/m_{\tilde{{G}}}^2$ while the two-photon
process does not depend on the gravitino mass. Therefore the search for
$\eeGGg$ at LEP is only sensitive if the gravitino is
ultralight ($m_{\tilde{{G}}}< 10^{-4}~\rm{eV/c^2}$).

If the gravitino is the LSP and it is the only supersymmetric particle
which is kinematically accessible, it is possible to search for
$\eeGG$~\cite{gravitino} by using ISR-photons to tag the undetectable
production of a gravitino pair.  This search makes it possible to set
a lower limit on the gravitino mass ($m_{\tilde{{G}}}$) and the
DELPHI collaboration is in this paper updating the calculation of this limit using
all available data with $\sqrt{s}>$180~GeV.

The NLSP lifetime depends on the gravitino mass and if this is
sufficiently large (a few hundred eV/c$^2$) the neutralinos in
$\eeGgGg$ will decay in such a way that the detected photons will not
originate at the beam interaction point.  When the mean decay paths
are comparable to the detector scale, events with a single photon not
pointing to the interaction region are expected and has been
searched for.

In gravity-mediated SUSY breaking (SUGRA) models~\cite{SUGRA}, the
gravitino is typically heavy with a mass of several hundreds of
GeV/c$^2$ and it is the neutralinos which are the LSP and NLSP.  Under
certain assumptions, $\tilde{\chi}^0_2$ will decay into a stable
undetectable $\tilde{\chi}^0_1$ and a photon.  This can give rise to
processes such as $\eeXXg$ and $\eeXgXg$ with both single- and
two-photon production.  A search for the two-photon process has
been carried out
but since the predicted cross-section for the single-photon process is
small the previously mentioned model-independent search for
$e^+e^-\rightarrow XY \rightarrow YY\gamma$ will be presented instead.

There exists in addition a so-called ``no-scale" supergravity 
model (also known as
the LNZ model)~\cite{LNZ} in which the local
supersymmetry breaking is decoupled from the global supersymmetry
breaking.  This results in the prediction of an ultralight gravitino
and the process $\eeGGg$.  The main phenomenological difference
between this model and the GMSB model is that only the gravitino and
neutralino masses are free parameters since the selectron masses and
the neutralino composition depend on the neutralino mass in the no-scale
supergravity model.

This paper describes the single- and multi-photon selection criteria,
the measurement of the single-photon cross-section and the number of
neutrino families and it gives limits on non-Standard Model physics
obtained from searches for gravitons in extra-dimensional models and
supersymmetric particles.

\section{The DELPHI detector}

The general criteria for the selection of events are based mainly on
the electromagnetic calorimeters and the tracking system of the DELPHI
detector~\cite{delphi_detec}.  All three major electromagnetic
calorimeters in DELPHI, the High density Projection Chamber (HPC), the
Forward ElectroMagnetic Calorimeter (FEMC) and the Small angle TIle
Calorimeter (STIC), have been used in the single-photon reconstruction
while only the FEMC and the HPC were used in the multi-photon analysis.
For the study of the non-pointing single-photon events only the HPC
detector was used.

The barrel region of DELPHI was covered by the HPC, which was a gas
sampling calorimeter able to sample a shower nine times
longitudinally.  The FEMC was made up of an array of 4532 lead glass
blocks in each endcap.  The energy resolution of this calorimeter was
degraded by the material in front of it, which caused photon
conversions and even preshowers.  The very forward luminosity monitor
STIC~\cite{stic} consisted of two cylindrical lead-scintillator
calorimeters read out by wavelength-shifting fibres.  Two layers of
scintillators mounted on the front of each STIC calorimeter together
with a smaller ring-shaped scintillator mounted directly on the
beampipe, provided $e - \gamma$ separation.  The angular coverages of
these calorimeters and the energy resolutions\footnotemark are given in
Table~\ref{calorimeters} and the detailed characteristics and
performances are described in~\cite{delphi_detec,stic}.

\footnotetext{The FEMC energy resolution in Table~\ref{calorimeters} 
was measured with electrons that had not interacted before reaching the
calorimeter and if preshowering electrons are taken into account the 
constant term has to be increased to 0.057 .}

Three different triggers were used in DELPHI to select single-photon
events.  The HPC trigger for purely neutral final states used a plane
of scintillators inserted into one of the HPC sampling gaps at a depth
of around 4.5 radiation lengths.  A second level trigger decision was
produced from the signals of analog electronics and was based on a
coincidence pattern inside the HPC module.  The trigger efficiency has
been measured with Compton and Bhabha events.  It was strongly
dependent on the photon energy, $E_\gamma$, rising steeply up to
$\sim$10~GeV, with about $52\%$ efficiency at 6~GeV and above $77\%$
when $E_{\gamma}>30$~GeV.  It reached a maximum of $84\%$ at
$E_{\gamma} \simeq E_{\rm beam}$. 
The FEMC trigger required an energy deposition of at least 2.5~GeV.  The
efficiency increased with energy and was 93\% at 10~GeV and above
99\% for $E_{\gamma}>15$~GeV.
Correlated noise in several adjacent channels caused fake triggers,
but these could be rejected offline with high efficiency by algorithms
that took into account the lead glass shower pattern.  The STIC
trigger required an energy deposition of at least 15~GeV and it was 
furthermore required that
there was no signal in at least one of the two scintillator planes in
front of the shower. The energy requirement caused inefficiencies for
showers with less than 30 GeV of energy. The efficiency of the scintillator
requirement has been
measured with samples of photons from $e^+e^-\gamma$ and
$q\bar{q}\gamma$ events.  The efficiency varied between 70\% and 33\%
over the angular region used in the analysis.

In addition to the electromagnetic calorimeters, the DELPHI tracking
system was used to reject events in which charged particles were
produced.  The main tracking devices were the Time Projection Chamber
(TPC) and the silicon microVertex Detector (VD) and its extension into
the forward region, the Very Forward Tracker (VFT).  The silicon
trackers were also used for electron/photon separation by vetoing
photon candidates which could be associated with hits in these
detectors.

Finally, the Hadron CALorimeter (HCAL) and its cathode-read-out system
were used to reject cosmic rays and to provide photon/hadron
separation, while the DELPHI Hermeticity Taggers were used to ensure
complete detector hermeticity for additional neutral particles.


\begin{table}[hbt]
\begin{center}
\begin{tabular}{|c||c|c|c|c|}
\hline
  & Type & Angular coverage & $\sigma_{E}/E$ & X$_0$  \\
\hline
\hline
 STIC: & Lead/scint. & \small{$2^\circ <\theta<10^\circ$~,~$170^\circ <\theta<178^\circ$} & 
\small{$0.0152\oplus(0.135/\sqrt{E})$}  & 27 \\
\hline
 FEMC: & Lead glass & \small{$10^\circ <\theta<37^\circ$~,~$143^\circ <\theta<170^\circ$} & 
\small{$0.03\oplus(0.12/\sqrt{E})\oplus(0.11/E)$} & 20 \\
\hline
 HPC: & Lead/gas & \small{$40^\circ <\theta < 140^\circ$} & 
\small{$0.043\oplus(0.32/\sqrt{E})$} & 18 \\
\hline
\end{tabular}
\end{center}
\caption{Polar angle coverage, energy resolution (where $E$ is in GeV
and $\oplus$ denotes addition in quadrature)
and thickness (in radiation lengths) of the electromagnetic calorimeters
in DELPHI.}
\label{calorimeters}
\end{table}

\section{Event selection}

\subsection{Single-photon events}

The single-photon events were selected in two stages.  In the first
stage, events with only one detected photon were preselected and compared to the
Standard Model process $\eenng$.  These events were also used to
calculate cross-sections and for a determination of the number of
light neutrino families.  A likelihood ratio method was then used to
maximize the sensitivity in the various single-photon searches.

\subsubsection{Preselection}

Different analyses were made depending on the 
electromagnetic calorimeter in which the photon candidate was found:

\begin{itemize}

\item 
Events with a photon in the HPC were selected by requiring a
shower having a scaled energy $x_{\gamma}=E_{\gamma}/E_{beam}>0.06$
and a polar angle, $\theta$, between $45^\circ$ and $135^\circ$ and no
charged particle tracks.  Electronic noise and alpha particles emitted
from the lead converter created fake low-energy showers in the HPC but
these could be rejected very effectively by requiring the longitudinal
shower profile to satisfy conditions defining a good electromagnetic
shape~\cite{delphi_130}.  Backgrounds from radiative Bhabha events and
Compton events were rejected by requiring no other electromagnetic
showers in the event unless they were in the HPC and within $20^\circ$
of the first one.  Cosmic rays were rejected mainly by the hadron
calorimeter.  If there were two or more hadronic showers the event was
discarded and if only one HCAL shower was present, the event was
rejected if the shower was not consistent with being caused by
punch-through of the electromagnetic shower.  The requirement of no
charged particles also removed cosmic ray background.
In addition, a constraint on the
$\gamma$ direction was imposed, requiring that the line of flight from
the mean interaction point and the shower direction measured in the
calorimeter coincided within $15^\circ$.  

The offline photon identification and reconstruction
efficiency was determined on the basis of a
Monte Carlo sample of events passed through the complete simulation of
the DELPHI detector~\cite{delphi_detec}. The efficiency 
depended on the photon energy and it ranged from 41\% at 6~GeV 
to 78\% for $E_{\gamma}>80$~GeV. In addition, some efficiencies 
were determined from data.  In particular,
the requirements of no electromagnetic or hadronic showers and no
charged particles were studied.  A sample of events triggered at
random and a sample of back-to-back Bhabha events with the electrons
in the STIC were used for this purpose.  It was found that noise and
machine background caused showers and tracks which would veto about
7\% of the good single-photon events and the data was corrected for this
loss.

\item 
Events with at least one shower in the FEMC with a scaled energy
$x_{\gamma}>0.10$ and a polar angle in the intervals
$12^\circ<\theta<32^\circ$ or $148^\circ<\theta<168^\circ $ were
selected.  The large background of radiative Bhabha events made it
necessary to add an energy-dependent angular requirement,
$\theta>28^\circ-80^\circ\cdot x_{\gamma}$ (and the complementary
$\theta$-region). This cut only affected the
very low energy region $x_{\gamma}<0.20$ .  
In order to separate
electrons from photons, the event was rejected if hits in the silicon
microvertex detectors (VD and VFT) could be associated with the
FEMC shower.  
The showers in the inner ($10^\circ<\theta<12^\circ$) 
and outer ($32^\circ<\theta<37^\circ$) radial parts of the FEMC
were discarded because of the large amount of material (about $2X_0$)
in front of the FEMC due to the STIC and the TPC detectors.  The
material in front of the FEMC meant that about half of the photons
preshowered before reaching the calorimeter.  Most of the preshower
was contained in a cone of about 15$^\circ$ around the largest shower
and the selection took this into account by requiring no charged
particle tracks, no other electromagnetic showers and no hadronic
showers outside a 15$^\circ$ cone.  If there were no charged particle
tracks inside the cone either, i.e., the photon had not preshowered,
it was required that only one FEMC shower was present in the event.
If, on the other hand, charged particle tracks were present in the
cone, more FEMC showers were allowed and their momentum vectors were
added to that of the largest shower.  The requirement of no
electromagnetic showers outside the cone greatly reduced the
background of radiative Bhabha and Compton events by rejecting events
that had one or both electrons in the acceptance of the experiment.
Events due to cosmic rays were rejected by the requirement of no
hadronic showers outside the cone.  Inside the cone, hadronic energy
was allowed only in the first layer of the HCAL.

Most reconstruction and event selection efficiencies in the analysis
were taken into account by using Monte Carlo samples passed through
the extensive detector simulation package of DELPHI~\cite{delphi_detec}. 
These efficiencies were estimated to be between 57\% and 75\% .
The additional loss of events due to noise and machine background was estimated
by using Bhabha events and random triggered data samples. 
The calculation showed that 
about 11\% of the single-photon events in FEMC were lost. 

\item 
Single photons in the STIC were preselected by requiring one
shower with a scaled energy $x_{\gamma}>0.30$ and with
$3.8^\circ<\theta<8^\circ$ or $172^\circ<\theta<176.2^\circ$.  The
large background coming from beam-gas interactions made it necessary
to add an energy-dependent angular requirement,
$\theta>9.2^\circ-9^\circ\cdot x_{\gamma}$ (and the complementary
$\theta$-region). This cut only affected the
energy region $x_{\gamma}<0.60$ .  It was furthermore required
that there were no other electromagnetic showers, no hadronic showers
and no charged particles in the event.  All single-photon candidates
had to satisfy the STIC single-photon trigger. A requirement of no 
signal in the small
scintillators mounted on the beampipe made it possible to reject
radiative $ee\gamma$ background.

The losses of photons in STIC, due to the offline photon identification 
and reconstruction, were estimated to be 5\% in an analysis using
$e^+e^-\gamma$ and $q\bar{q}\gamma$ events.  The selection
of events with no additional shower in the STIC and no tracks implied similar
losses (7\%) to those found in the HPC analysis and were estimated with the
same methods.

\end{itemize}

The result of the single-photon event selection described above was that events 
with more than one photon could survive if the other photons were at low angle
($\theta_{\gamma} < 2.2^\circ$), low energy ($E_{\gamma} <$ 0.8 GeV)
or within 3$^\circ$, 15$^\circ$ and 20$^\circ$ from the highest energy
photon in the STIC, FEMC and HPC respectively.

\subsubsection{Final selection}

A likelihood ratio method was used to select the final sample of
events in the various searches using single-photon events.  The photon
energy was used as the final discriminating variable and two
likelihood functions ($f_S(E_{\gamma})$ and $f_B(E_{\gamma})$) were
produced from the normalized photon energy distributions of the
simulated signal and background events, after passing through the same
selection criteria.  The likelihood ratio function was defined as
${\cal L}_R = f_S(E_{\gamma})/f_B(E_{\gamma})$ where an event with
${\cal L}_R > {\cal L}_{R}^{CUT}$ was selected as a candidate event.
The value of ${\cal L}_{R}^{CUT}$ was optimised to give the minimum
signal cross-section excluded at 95\% C.L.  in the absence of a
signal:  $$\sigma^{min}({\cal L}_{R}^{CUT}) = \frac{N^{min}_{95}({\cal
L}_{R}^{CUT})} {\epsilon^{max}({\cal L}_{R}^{CUT}) \times L}$$ where
$N^{min}_{95}$ is the upper limit on the number of signal events at
95\% C.L., $\epsilon^{max}$ is the efficiency for the signal and L is
the integrated luminosity.  This method optimises the background
suppression for a given signal efficiency~\cite{lrref}.  A more
detailed description of the method can be found in~\cite{cha189}.

The data collected at different centre-of-mass energies were
analysed separately and different analyses were made depending on the 
electromagnetic calorimeter in which the photon was recorded.
The final experimental limits were obtained using a 
Bayesian multi-channel method~\cite{cha189} which combined the results
from up to 20 analyses. The method 
takes into account all the available information (such as the fraction 
of the signal and 
the average background in each subdetector and in each data subsample), 
and this makes it possible to calculate optimum limits.

\subsection{Multi-photon events}

Multi-photon events were selected by a two-step procedure.  In the
first preselection step, events with at least two photons and missing energy
were selected.  This sample was dominated by the Standard Model
process $\eenngg$ and it was used to monitor the modelling of this
process by the KORALZ 4.02
generator~\cite{koralz}.  In a second step, the selection criteria
were tightened in order to improve the experimental sensitivity to
possible signals of supersymmetry, such as the $\eeGgGg$ or $\eeXgXg$
processes.  This was achieved by imposing more stringent requirements
on the photon polar angles as well as on the event missing mass and
transverse momentum.

\subsubsection{Preselection}

The preselection required that there were at least two electromagnetic
clusters in the HPC and FEMC detectors with $x_{\gamma} > 0.05$ and
that the event had a total visible energy $E_{vis}<0.9\sqrt{s}$ and a
total transverse momentum $p_T > 0.03E_{miss}$ where
$E_{miss}=\sqrt{s}- E_{vis}$.  The polar angles of the clusters were
required to be in the region $10^\circ<\theta<170^\circ$ with at least
one cluster in $25^\circ<\theta<155^\circ$.  The clusters were
identified as photons by requiring that there were no reconstructed
charged particle tracks and no hits in the VD or VFT detectors that
could be associated with the clusters.

All FEMC and HPC showers in a cone of $10^\circ$ around the
shower with the highest energy 
were merged together.  Noise and clusters from alpha particles
in the HPC were rejected by requirements on the longitudinal shower
profile~\cite{delphi_gg} and the reconstructed HPC shower axis had to
point to the interaction point within $25^\circ$.

In order to reduce the background from $\eegg$ events it was required
that the acoplanarity was greater than $3^\circ$.  
A requirement of no
hits in the Hermeticity Taggers reduced the background of
$e^+e^-\rightarrow \gamma \gamma\gamma$ where one photon was lost in
the holes between the calorimeters.  By requiring that the
acoplanarity was less than $140^\circ$ when $|\theta_{\gamma
1}-\theta_{\gamma 2}|<20^\circ$ it was possible to eliminate single-photon 
events with the photon converting in the material in front of
the calorimeters and producing two separate clusters.  Cosmic ray events
were rejected by vetoing on energy deposits in the external layers of
the HCAL.  Doubly radiative Bhabha events where the electrons escape
detection along the beampipe could be rejected by requiring that the
energy in the STIC was less than $0.02\sqrt{s}$ and that the polar angle
of the missing momentum was in the region
$10^\circ<\theta_{miss}<170^\circ$.

If an event had three electromagnetic clusters passing the
previous selection it was retained if it had
$E_{vis}<0.8\sqrt{s}$ and the sum of the angles between the three observed
photons was less than $358^\circ$, i.e., if the event was
significantly aplanar.  Events with four or more
electromagnetic clusters were discarded.

\subsubsection{Final selections}

The final selection in the search for $\eeGgGg$ required that the
scaled transverse momentum for each cluster satisfied
$p_{T\gamma}/E_{beam}>0.07$ if the missing mass ($M_{miss}$) was
larger than 60 GeV/c$^2$. This requirement strongly suppressed the
Standard Model background coming from $\ee \to \nu \bar{\nu} \gamma \gamma
(\gamma)$ where the photons, being emitted from the initial state
particles, have a relatively low transverse momentum.  A more
stringent requirement of $p_{T\gamma}/E_{beam} > 0.14$ was used when 
the missing mass was close to the
Z mass ($80<M_{miss}<110$~GeV/c$^2$) and the background was larger.  
It was in addition required
that the energies and angles of the detected photons were compatible
with those expected for a particular $\tilde{\chi}^0_1$ mass.

In the search for $\eeXgXg$, the cut on transverse momentum was
replaced by a cut on the polar angle of the photons.  It was
required that $20^\circ<\theta_{\gamma}<160^\circ$ for
$M_{miss}>60$~GeV/c$^2$ with the exception of the Z-region 
($80< M_{miss}<110$~GeV/c$^2$) where the polar angle was required to 
fulfil
$40^\circ<\theta_{\gamma}<140^\circ$. The energies of the detected 
photons
had to be compatible with those expected for a particular pair of
$\tilde{\chi}^0_1$ and $\tilde{\chi}^0_2$ masses.

The efficiency of the selection was studied using simulated $\eeGgGg$
and $\eeXgXg$ events.  The estimated efficiency was in the range
40-55\% for both search scenarios~\cite{delphi_gg}.

\subsection{Non-pointing single-photon events}

The fine granularity of the HPC calorimeter provided a precise
reconstruction of the axis direction in electromagnetic showers.  This
feature was used to select events with a single photon whose flight
direction did not point to the beam interaction region.  Events with a
single non-pointing photon are expected when two neutral particles
with large mean decay paths ($> 4$ m) are produced which subsequently
decay into a photon and an invisible particle since the probability
that both photons are recorded is then small.

Events of this kind were searched for by requiring one photon in the
HPC calorimeter with $E_\gamma>10$~GeV and an impact parameter
exceeding 40 cm.  A more stringent HPC cluster selection was made
compared to the multi-photon selection~\cite{delphi_gg}.  This was
done in order to reject noise and cosmic ray events which were a larger
problem in the non-pointing single-photon analysis.

It was required that the events had no reconstructed tracks and that
no hits in the vertex detector could be found along a line going
from the interaction point to the electromagnetic shower.  
Cosmic ray events, which represented the main
background, were largely reduced by vetoing on isolated
hits or tracks in the Hermeticity Taggers and signals from the
external HCAL layers and from the cathode-read-out system of the
hadron calorimeter.

The efficiency of the selection was studied using simulated $\eeGgGg$
events that had passed through the entire DELPHI detector simulation
package.  For a mean decay path of 3-8 m the probability that a neutralino would decay
before the HPC was small and the total efficiency (including acceptance)
was only 4-5\%~\cite{delphi_gg}.

\section{Real data and simulated samples}

\subsection{Data samples}

The ten data samples that were used in the single-photon analysis are
summarized in Table~\ref{lumi}.  These samples were recorded by DELPHI
during 1997 to 2000.  The multi-photon analysis also included data
recorded during 1995 and 1996 at the centre-of-mass energies of 130,
136, 161 and 172 GeV.  These samples added another 21 pb$^{-1}$ of
integrated luminosity.

\begin{table}[hbt]
\begin{center}
\begin{tabular}{|c|c|c|c|c|c|}
\hline
\multicolumn{1}{|c|}{Year} &
\multicolumn{2}{|c|}{$\sqrt{s}$ (GeV)} &
\multicolumn{3}{|c|}{Luminosity~(pb$^{-1}$)} \\
\cline{2-6}
\multicolumn{1}{|c|}{  } & \multicolumn{1}{|c|}{bin} & \multicolumn{1}{|c|}{average} & 
\multicolumn{1}{|c|}{ HPC }  & \multicolumn{1}{|c|}{ FEMC}  & \multicolumn{1}{|c|}{ STIC } \\
\hline
\hline
\multicolumn{1}{|c|}{1997} & \multicolumn{1}{|c|}{180.8-184.0} & \multicolumn{1}{|c|}{182.7} &
\multicolumn{1}{|c|}{50.2} & \multicolumn{1}{|c|}{49.2} & \multicolumn{1}{|c|}{51.4} \\

\multicolumn{1}{|c|}{1998} & \multicolumn{1}{|c|}{188.3-189.2} & \multicolumn{1}{|c|}{188.6} &
\multicolumn{1}{|c|}{154.7} & \multicolumn{1}{|c|}{157.7} & \multicolumn{1}{|c|}{157.3} \\

\multicolumn{1}{|c|}{1999} & \multicolumn{1}{|c|}{191.4-191.8} & \multicolumn{1}{|c|}{191.6} &
\multicolumn{1}{|c|}{25.9} & \multicolumn{1}{|c|}{25.9} & \multicolumn{1}{|c|}{25.9} \\

\multicolumn{1}{|c|}{1999} & \multicolumn{1}{|c|}{195.4-195.9} & \multicolumn{1}{|c|}{195.5} &
\multicolumn{1}{|c|}{76.4} & \multicolumn{1}{|c|}{76.4} & \multicolumn{1}{|c|}{76.4} \\

\multicolumn{1}{|c|}{1999} & \multicolumn{1}{|c|}{199.1-200.0} & \multicolumn{1}{|c|}{199.5} &
\multicolumn{1}{|c|}{83.4} & \multicolumn{1}{|c|}{83.4} & \multicolumn{1}{|c|}{83.4} \\

\multicolumn{1}{|c|}{1999} & \multicolumn{1}{|c|}{201.4-202.0} & \multicolumn{1}{|c|}{201.6} &
\multicolumn{1}{|c|}{40.6} & \multicolumn{1}{|c|}{40.6} & \multicolumn{1}{|c|}{40.6} \\

\multicolumn{1}{|c|}{2000} & \multicolumn{1}{|c|}{202.0-204.5} & \multicolumn{1}{|c|}{203.7} &
\multicolumn{1}{|c|}{8.4} & \multicolumn{1}{|c|}{8.4} & \multicolumn{1}{|c|}{8.4} \\

\multicolumn{1}{|c|}{2000} & \multicolumn{1}{|c|}{204.5-206.0} & \multicolumn{1}{|c|}{205.2} &
\multicolumn{1}{|c|}{76.2} & \multicolumn{1}{|c|}{76.3} & \multicolumn{1}{|c|}{76.1} \\

\multicolumn{1}{|c|}{2000} & \multicolumn{1}{|c|}{206.0-207.5} & \multicolumn{1}{|c|}{206.7} &
\multicolumn{1}{|c|}{121.6} & \multicolumn{1}{|c|}{125.7} & \multicolumn{1}{|c|}{125.6} \\

\multicolumn{1}{|c|}{2000} & \multicolumn{1}{|c|}{207.5-209.2} & \multicolumn{1}{|c|}{208.2} &
\multicolumn{1}{|c|}{8.3} & \multicolumn{1}{|c|}{8.4} & \multicolumn{1}{|c|}{8.4} \\

\hline
\end{tabular}
\end{center} 
\caption{The different datasets used in the single-photon analysis.
}
\label{lumi} 
\end{table}

\subsection{Simulation of the Standard Model background}

\subsubsection{The single-photon analysis}

Apart from the $\eenng (\gamma)$ process, single-photon events can be
faked by the QED reaction $e^+e^- \rightarrow e^+e^-\gamma $ if the
two electrons escape undetected along the beampipe or if the electrons
are in the detector acceptance but are not detected by the experiment.

This process has a very high cross-section, decreasing rapidly when
the energy ($E_{\gamma}$) and the polar angle ($\theta_{\gamma}$) of
the photon increase.  The behaviour of this QED background together
with the rapidly varying efficiencies at low energies are the reasons
why different energy cuts had to be applied to photons in the three
calorimeters:  $x_{\gamma}>0.06$ (HPC), $x_{\gamma}>0.10$ (FEMC) and
$x_{\gamma}>0.30$ (STIC).  The energy-dependent cut on the polar angle
in the FEMC and STIC analyses were also necessary to reduce this
background.  Another critical parameter in the rejection was the polar
angle at which the electrons start being seen in the STIC detector.
This detector reconstructs electrons down to $\theta =2.2^\circ$ and
in addition, the scintillator counters mounted on the beampipe could
be used to reject events with electrons down to $1.8^\circ$.
Simulations have shown that even at lower angles (down to
$0.97^\circ$) a large fraction of the electrons were detectable
because they interacted with a tungsten shield mounted inside the
beampipe and leaked enough energy into the STIC to make it possible to
reject the events.

The remaining background from the $e^+e^-\gamma$ process was
calculated with a Monte Carlo program~\cite{TEEGG} and two different
event topologies were found to contribute.  
Either both electrons were below the
STIC acceptance or one of the electrons was in the DELPHI acceptance
where it was wrongly identified as a photon, while the photon was lost
in the holes between the electromagnetic calorimeters.  The first
topology gives background at low photon energy while the second one
produces fake photon events at high energy.

In the STIC analysis, an additional background was due to the single
electrons produced by interactions between the beam particles and
residual gas molecules in the LEP beampipe.  In these $e\rightarrow
e\gamma$ events, the photons were always lost in the beampipe while the
off-energy electrons were bent into the STIC acceptance by the
low-beta quadrupoles close to the DELPHI experiment.  The rate of this 
background was
so large that it was not possible to provide a $\gamma-e$ separation
powerful enough to eliminate it completely.  A simulation
has been made of off-energy electron production~\cite{off}, but it
could not be used in the analysis since the vacuum pressure around the
LEP ring was not known to the required precision.  Instead, a
background sample was collected with a trigger similar to the photon
trigger except that it did not use the scintillators for
photon-electron separation.  After applying all the cuts used in the
single-photon analysis, except the scintillator requirements, this
background sample was used to estimate the remaining off-energy
electron background and 19\% of the overall background was estimated
to be caused by off-energy electrons.

The contributions from other processes such as $\gamma\gamma$
collisions~\cite{2gamma}, $e^+e^- \rightarrow \gamma \gamma (\gamma)$~\cite{radcor}, 
cosmic ray events, $e^+e^- \rightarrow \mu^+ \mu^- \gamma$~\cite{koralz}, $e^+e^-
\rightarrow \tau^+ \tau^- \gamma$~\cite{koralz} and four-fermion events~\cite{4fermion} 
have also been calculated.

The $\nu\bar{\nu}\gamma(\gamma)$ process was simulated by the
KORALZ~\cite{koralz} program. Comparisons of the predicted 
cross-section by KORALZ 4.02 has been made with the cross-section 
predicted by the NUNUGPV~\cite{nunugpv} and KK 4.19~\cite{kk}
programs. The total cross-section in the STIC, FEMC and HPC acceptance
was calculated to be 1.0\% lower by the KK program than by the KORALZ
program. In the acceptance relevant to most searches, i.e., 
for events with a photon with $x_{\gamma}<0.70$ in the FEMC or the HPC,
KK predicted a cross-section that was 1\% lower than the prediction
by KORALZ. The NUNUGPV program has earlier been shown to give
predictions within 1\% of those of the KORALZ program~\cite{delphi_189}
in measurements using the full energy range of the photons. 
These theoretical uncertainties are insignificant compared
to the statistical and systematic uncertainties in the 
present measurement.

\subsubsection{The multi-photon analysis}

Multi-photon final states can be produced at LEP via the reactions
$\eenngg$ and $e^+e^- \rightarrow \gamma\gamma(\gamma)$.  In the case
of double ISR with final state $\nu\bar{\nu}$ production, the photons
tend to have a relatively low transverse momentum and tend to have
both large acoplanarity and large acollinearity.  Since the neutrino
production is mainly mediated by Z-exchange in the s-channel, the
missing mass distribution has a large peak corresponding to the mass
of the Z.  The $\eenngg$ process was, as in the single-photon case,
simulated by the KORALZ generator~\cite{koralz}.

The process $e^+e^- \rightarrow \gamma\gamma(\gamma)$ is a QED
interaction between the incoming electron and positron and it is 
mediated by an
electron in the t-channel.  The RADCOR model~\cite{radcor} was used to
simulate this background.  The process is easy to reject when only two
hard photons are produced but if additional hard ISR photons are
emitted and lost in the beampipe, the visible photon pair can show
relatively large acollinearity but small acoplanarity.

Additional minor background contributions can come from double
radiative Bhabha events $e^+e^- \rightarrow e^+e^-\gamma\gamma$ when
both electrons remain undetected or from radiative Bhabha events
where one or more electrons are wrongly identified
as photons.
The BHWIDE program~\cite{bhwide} was used as an event
generator for these events.  A requirement of a minimum transverse
missing momentum very effectively eliminated the cases where both
electrons escaped along the beampipe.

\section{Comparison with Standard Model expectations}

\subsection{Single-photon cross-section}

\begin{figure}[htb]
\centerline{\epsfig{file=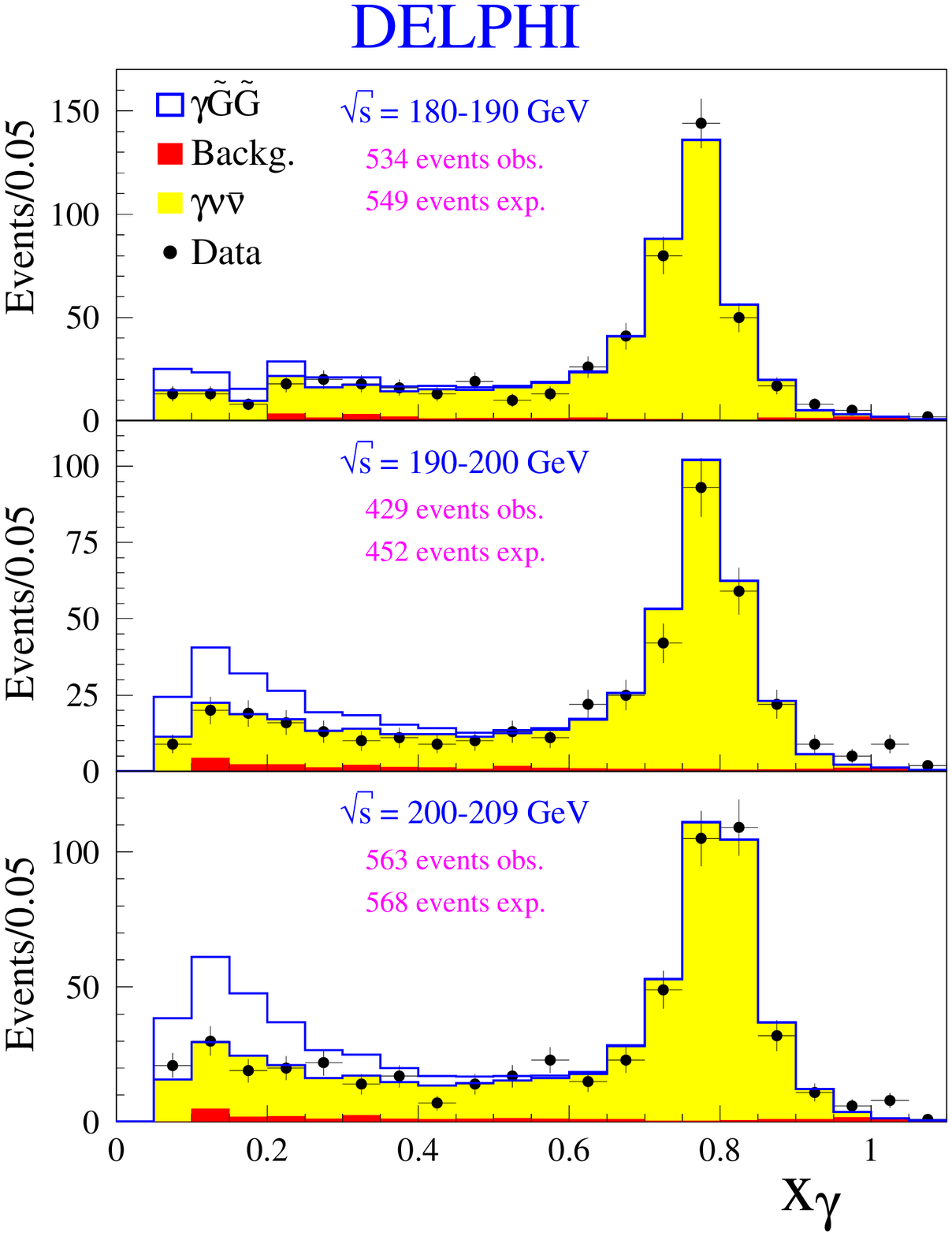,
width=15cm}}
\caption[]{
$x_{\gamma}$ of selected single photons for three $\sqrt{s}$-bins.
The light shaded area is the expected distribution from
$e^+e^- \rightarrow \nu \bar{\nu}\gamma$ and the
dark shaded area is the total background from other
sources. Indicated in the plot is also the signal expected
from $\eeGG$ under the assumption that 
$m_{\tilde{{G}}}= 7 \cdot 10^{-6} ~ \rm{eV/c^2}$.
}
\label{somma_xgam} 
\end{figure}

The energy spectrum of the 1526 selected single-photon events from all
calorimeters is shown in Figure~\ref{somma_xgam} together with the
expected contributions from known sources.  The $x_{\gamma}$
distributions are shown for three $\sqrt{s}$-bins and the integrated luminosity
and average $\sqrt{s}$ of the datasets that make up these bins are
given in Table~\ref{lumi}.  The $\nu\bar{\nu}\gamma$ process was
simulated by the KORALZ~\cite{koralz} program and then passed through
the extensive detector simulation package of DELPHI~\cite{delphi_detec}.

\begin{figure}[htb]
\centerline{\epsfig{file=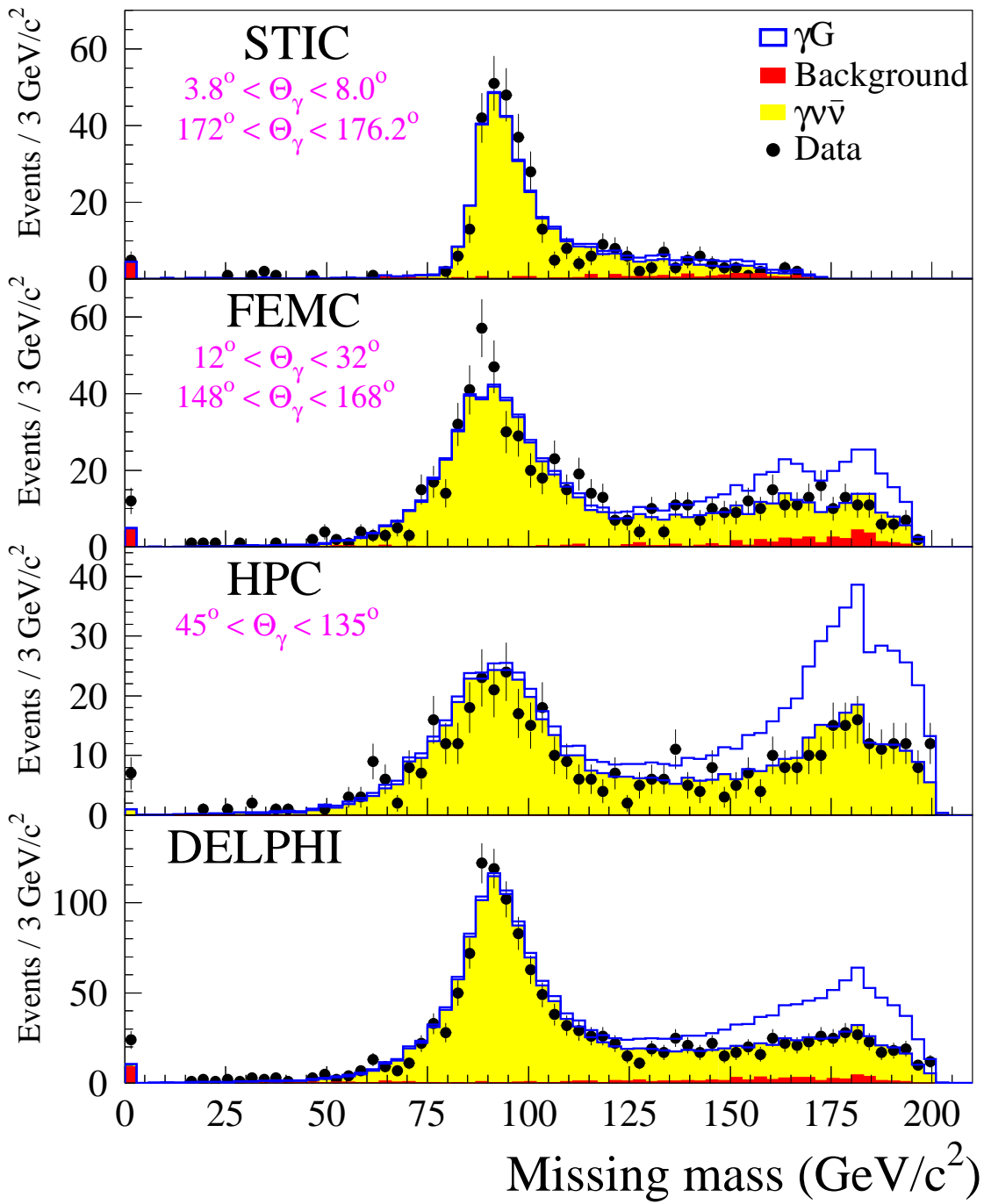,width=15cm}}
\caption[]{
Missing mass (or recoil mass) distributions of all single-photon events
in DELPHI (from data recorded at $\sqrt{s}$ = 180-209 GeV).
The figure shows the missing mass distribution from each 
calorimeter separately and the bottom plot shows the combined spectrum.
The light shaded area is the expected distribution from
$e^+e^- \rightarrow \nu \bar{\nu}\gamma$ and the
dark shaded area is the total background from other
sources. The signal expected from $e^+e^-\rightarrow\gamma G$ production
is indicated (two extra dimensions and $M_{D}=0.75$~TeV/c$^2$ were assumed
in this calculation).
}
\label{missing_mass} 
\end{figure}

The missing mass (or recoil mass) distributions of the events recorded
at $\sqrt{s}$=180-209~GeV are shown in Figure~\ref{missing_mass}.  In
this plot the distributions obtained with each of the three DELPHI
electromagnetic calorimeters are shown separately. Data samples
recorded at different $\sqrt{s}$ extend to different maximum values
of missing mass and this creates ``steps'' in the distributions. 
Events where the measured energy of the photon is larger than the collision 
energy have been put in the first bin. The number of these events is 
larger in the data than expected. The simulation of background
sources show that some of these events can be explained by 
QED two-photon events with one of the photons escaping detection.
It is possible that the excess is due to temporary problems with
some calorimeter channels that were not perfectly reproduced in the
simulation and which created a larger QED background in the data sample. 
Otherwise there is no sign of an excess above the 
Standard Model expectation in any of the three calorimeter analyses 
nor in any of the $\sqrt{s}$-bins.

The number of events and cross-sections obtained from the event
samples after correcting for background and efficiencies are given in
Table~\ref{cross}.  This table does not include the FEMC events with
$x_{\gamma}<0.20$ since they were used only in the searches and not in
the cross-section calculations.  The previously mentioned Monte Carlo
programs were used to calculate the expected values of the
cross-section of the process $e^+e^- \rightarrow \nu
\bar{\nu}\gamma(\gamma)$ inside the acceptance of each of the three
detectors used in the analysis.  The contributions from various
sources to the systematic error in the cross-section measurement are
given in Table~\ref{syst}.  The dominant uncertainty comes from the
estimation of trigger and detection efficiencies while the theoretical
uncertainty is insignificant compared with the experimental systematic 
errors.  

\begin{figure}[htb]
\centerline{\epsfig{file=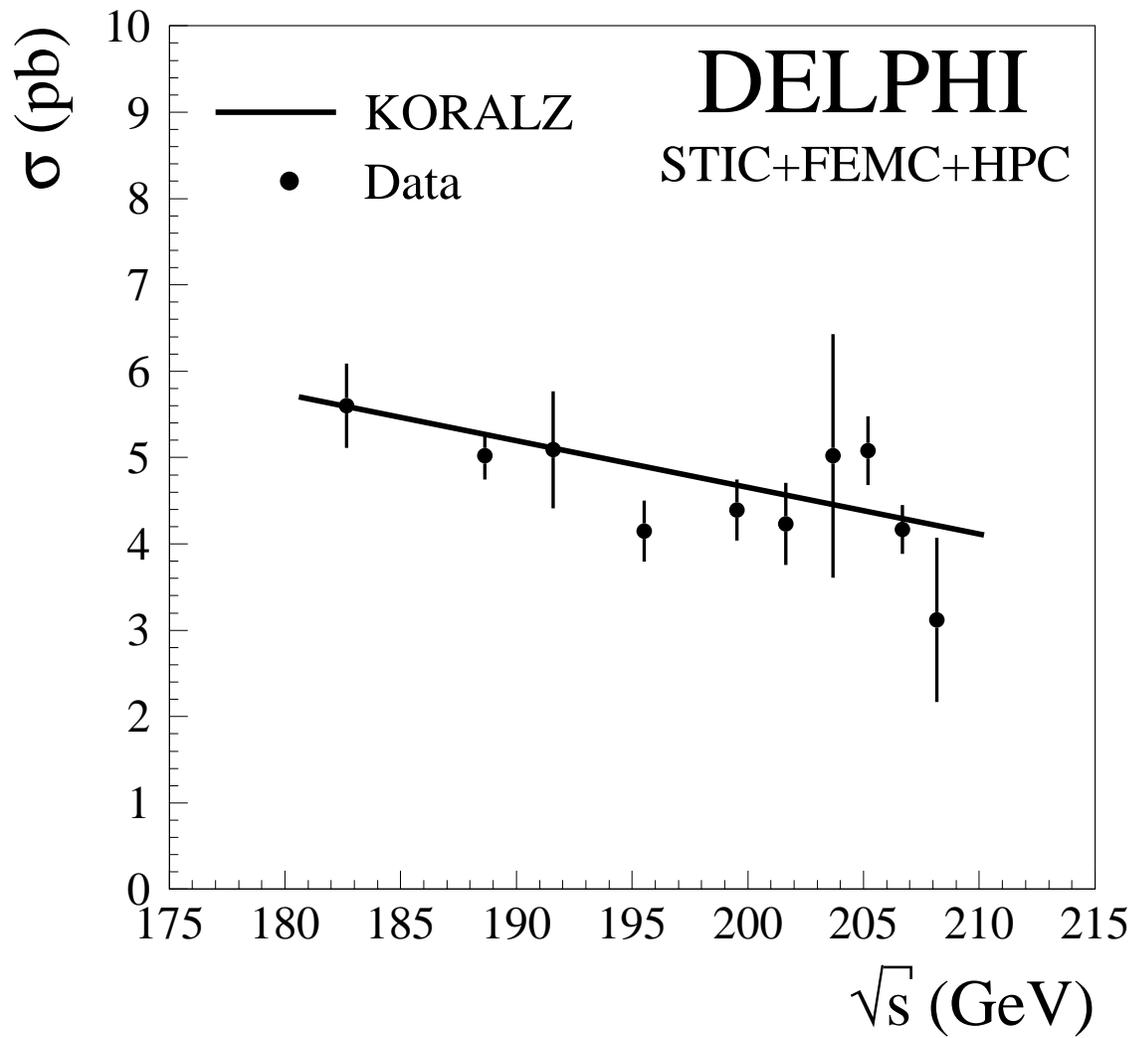,
width=15cm}}
\caption[]{
The single-photon cross-section measured by the STIC, FEMC and HPC detectors compared to the
cross-section predicted by Koralz.}
\label{sigma6}
\end{figure}

Figure~\ref{sigma6} shows the expected behaviour of the Standard Model
single-photon cross-section as a function of the LEP energy, compared
with the values measured with the STIC, FEMC and HPC calorimeters.  The
Standard Model
prediction, calculated by KORALZ, tends to be above the data but the 
difference is not statistically significant.

\begin{table}[hbt]
\begin{center}
\begin{tabular}{|c|c||c|c|c|}
\cline{2-5}
\multicolumn{1}{c}{ } & 
\multicolumn{1}{|c||}{$\sqrt{s}$ }   &
\multicolumn{1}{|c|}{ 180-190~GeV } & 
\multicolumn{1}{|c|}{ 190-200~GeV } &
\multicolumn{1}{|c|}{ 200-209~GeV } \\
\cline{2-5}
\multicolumn{1}{c}{ } & 
\multicolumn{1}{|c||}{$<\sqrt{s}>$ } &
\multicolumn{1}{|c|}{ 187.1~GeV }   & 
\multicolumn{1}{|c|}{ 196.8~GeV }   &
\multicolumn{1}{|c|}{ 205.4~GeV }   \\
\cline{2-5}
\hline
\multicolumn{1}{|c|}{ } & 
\multicolumn{1}{|c||}{ $N_{observed}$ } &
\multicolumn{1}{|c|}{ 126}         & 
\multicolumn{1}{|c|}{ 90}         &
\multicolumn{1}{|c|}{ 114}         \\
\cline{2-5}
\multicolumn{1}{|c|}{ STIC } & \multicolumn{1}{|c||}{$N_{background}$ } &
\multicolumn{1}{|c|}{ 10.1 } & 
\multicolumn{1}{|c|}{ 7.2 } &
\multicolumn{1}{|c|}{ 7.1 } \\
\cline{2-5}
\multicolumn{1}{|c|}{ $0.3<x_{\gamma}<0.9$ } & 
\multicolumn{1}{|c||}{ $N_{e^+e^-\rightarrow\nu\bar{\nu}\gamma}$ }      & 
\multicolumn{1}{|c|}{ 123.8$\pm$1.2 }   & 
\multicolumn{1}{|c|}{ 86.9$\pm$1.0 }   &
\multicolumn{1}{|c|}{ 112.9$\pm$1.7 }   \\
\cline{2-5}      
\multicolumn{1}{|c|}{ $3.8^\circ<\theta_{\gamma}<8^\circ$} & \multicolumn{1}{|c||}{ $\sigma_{meas}$~(pb) } &
\multicolumn{1}{|c|}{ 1.37$\pm$0.14 } & 
\multicolumn{1}{|c|}{ 1.22$\pm$0.14 } &
\multicolumn{1}{|c|}{ 1.12$\pm$0.11 } \\
\cline{2-5}
\multicolumn{1}{|c|}{ $172^\circ<\theta_{\gamma}<176.2^\circ$ } & \multicolumn{1}{|c||}{$\sigma_{\nu\bar{\nu}\gamma(\gamma)}$~(pb) } &
\multicolumn{1}{|c|}{ 1.44 }        & 
\multicolumn{1}{|c|}{ 1.29 }        &
\multicolumn{1}{|c|}{ 1.18 }        \\
\cline{2-5}
\multicolumn{1}{|c|}{  } & 
\multicolumn{1}{|c||}{ $N_{\nu}$ }           &
\multicolumn{1}{|c|}{ 2.83$\pm$0.31 }        & 
\multicolumn{1}{|c|}{ 2.81$\pm$0.37 }        &
\multicolumn{1}{|c|}{ 2.81$\pm$0.33 }        \\
\hline
\hline 
\multicolumn{1}{|c|}{ } & 
\multicolumn{1}{|c||}{ $N_{observed}$ } &
\multicolumn{1}{|c|}{ 220}         & 
\multicolumn{1}{|c|}{ 175}         &
\multicolumn{1}{|c|}{ 224}         \\
\cline{2-5}
\multicolumn{1}{|c|}{ FEMC } & \multicolumn{1}{|c||}{$N_{background}$ } &
\multicolumn{1}{|c|}{ 9.5 } & 
\multicolumn{1}{|c|}{ 8.9 } &
\multicolumn{1}{|c|}{ 9.0 } \\
\cline{2-5}
\multicolumn{1}{|c|}{ $0.2<x_{\gamma}<0.9$ } & 
\multicolumn{1}{|c||}{ $N_{e^+e^-\rightarrow\nu\bar{\nu}\gamma}$ }      & 
\multicolumn{1}{|c|}{ 208.3$\pm$2.2 }   & 
\multicolumn{1}{|c|}{ 168.1$\pm$1.1 }   &
\multicolumn{1}{|c|}{ 200.4$\pm$1.6 }   \\
\cline{2-5}      
\multicolumn{1}{|c|}{$12^\circ <\theta_{\gamma}< 32^\circ$  } & 
\multicolumn{1}{|c||}{ $\sigma_{meas}$~(pb)} &
\multicolumn{1}{|c|}{ 1.98$\pm$0.14 }        & 
\multicolumn{1}{|c|}{ 1.71$\pm$0.14 }        &
\multicolumn{1}{|c|}{ 1.71$\pm$0.12 }         \\
\cline{2-5}
\multicolumn{1}{|c|}{ $148^\circ <\theta_{\gamma}< 168^\circ$ } & 
\multicolumn{1}{|c||}{ $\sigma_{\nu\bar{\nu}\gamma(\gamma)}$~(pb)} &
\multicolumn{1}{|c|}{ 1.97 }   & 
\multicolumn{1}{|c|}{ 1.76 }   &
\multicolumn{1}{|c|}{ 1.57 }   \\
\cline{2-5}
\multicolumn{1}{|c|}{ } & 
\multicolumn{1}{|c||}{ $N_{\nu}$ }           &
\multicolumn{1}{|c|}{ 3.03$\pm$0.25 }        & 
\multicolumn{1}{|c|}{ 2.90$\pm$0.28 }        &
\multicolumn{1}{|c|}{ 3.30$\pm$0.27 }        \\
\hline
\hline 
\multicolumn{1}{|c|}{ } & 
\multicolumn{1}{|c||}{ $N_{observed}$ } &
\multicolumn{1}{|c|}{ 177}         & 
\multicolumn{1}{|c|}{ 127}         &
\multicolumn{1}{|c|}{ 190}         \\
\cline{2-5}
\multicolumn{1}{|c|}{ HPC } & \multicolumn{1}{|c||}{$N_{background}$ } &
\multicolumn{1}{|c|}{ 0.3 } & 
\multicolumn{1}{|c|}{ 0.2 } &
\multicolumn{1}{|c|}{ 0.1 } \\
\cline{2-5}
\multicolumn{1}{|c|}{ $0.06<x_{\gamma}<1.1$ } & 
\multicolumn{1}{|c||}{ $N_{e^+e^-\rightarrow\nu\bar{\nu}\gamma}$ }      & 
\multicolumn{1}{|c|}{ 190.1$\pm$2.6 }   & 
\multicolumn{1}{|c|}{ 151.4$\pm$1.1 }   &
\multicolumn{1}{|c|}{ 198.1$\pm$2.0 }   \\
\cline{2-5}      
\multicolumn{1}{|c|}{ $45^\circ <\theta_{\gamma}< 135^\circ$ } & 
\multicolumn{1}{|c||}{$\sigma_{meas}$~(pb)  } &
\multicolumn{1}{|c|}{ 1.78$\pm$0.13 }        & 
\multicolumn{1}{|c|}{ 1.41$\pm$0.13 }         &
\multicolumn{1}{|c|}{ 1.50$\pm$0.11 }        \\
\cline{2-5}
\multicolumn{1}{|c|}{ } & 
\multicolumn{1}{|c||}{  $\sigma_{\nu\bar{\nu}\gamma(\gamma)}$~(pb)} &
\multicolumn{1}{|c|}{ 1.89 }        & 
\multicolumn{1}{|c|}{ 1.75 }        &
\multicolumn{1}{|c|}{ 1.61 }        \\
\cline{2-5}
\multicolumn{1}{|c|}{  } & 
\multicolumn{1}{|c||}{ $N_{\nu}$ }           &
\multicolumn{1}{|c|}{ 2.76$\pm$0.29 }        & 
\multicolumn{1}{|c|}{ 2.18$\pm$0.30 }        &
\multicolumn{1}{|c|}{ 2.71$\pm$0.30 }        \\
\hline

\end{tabular}
\end{center}
\caption{
The number of selected and expected single-photon events and the
measured and calculated cross-sections with three DELPHI calorimeters for 
$e^+e^- \rightarrow \nu \bar{\nu}\gamma(\gamma)$
(KORALZ with three neutrino generations) and the number of neutrino generations
calculated from the cross-sections.
The errors are statistical only. 
}
\label{cross}
\end{table}

\begin{table}[hbt]
\begin{center}
\begin{tabular}{|c||c|c|c|c|c|c|}
\cline{2-7}
\multicolumn{1}{c||}{        } & \multicolumn{2}{|c|}{ HPC } & \multicolumn{2}{|c|}{ FEMC} & \multicolumn{2}{|c|}{ STIC } \\
\hline
\multicolumn{1}{|c||}{ Source } & 
\multicolumn{1}{|c|}{ Variation }  & \multicolumn{1}{|c|}{ $\Delta\sigma$ } &
\multicolumn{1}{|c|}{ Variation }  & \multicolumn{1}{|c|}{ $\Delta\sigma$ } &
\multicolumn{1}{|c|}{ Variation }  & \multicolumn{1}{|c|}{ $\Delta\sigma$ } \\
\hline
\multicolumn{1}{|l||}{ Luminosity } & 
\multicolumn{1}{|c|}{ $\pm$0.6\% }  & \multicolumn{1}{|c|}{ $\pm$0.6\% } &
\multicolumn{1}{|c|}{ $\pm$0.6\% }  & \multicolumn{1}{|c|}{ $\pm$0.6\% } &
\multicolumn{1}{|c|}{ $\pm$0.6\% }  & \multicolumn{1}{|c|}{ $\pm$0.6\% } \\
\multicolumn{1}{|l||}{ Trigger efficiency } & 
\multicolumn{1}{|c|}{ $\pm$5\% }  & \multicolumn{1}{|c|}{ $\pm$5\% } &
\multicolumn{1}{|c|}{ $\pm$2\% }  & \multicolumn{1}{|c|}{ $\pm$2\% } &
\multicolumn{1}{|c|}{ $\pm$6\% }  & \multicolumn{1}{|c|}{ $\pm$6\% } \\
\multicolumn{1}{|l||}{ Identification efficiency } & 
\multicolumn{1}{|c|}{ $\pm$5\% }  & \multicolumn{1}{|c|}{ $\pm$5\% } &
\multicolumn{1}{|c|}{ $\pm$6\% }  & \multicolumn{1}{|c|}{ $\pm$6\% } &
\multicolumn{1}{|c|}{ $\pm$5\% }  & \multicolumn{1}{|c|}{ $\pm$5\% } \\
\multicolumn{1}{|l||}{ Calorimeter energy scale } & 
\multicolumn{1}{|c|}{ $\pm$5\% }  & \multicolumn{1}{|c|}{ $\pm$4\% } &
\multicolumn{1}{|c|}{ $\pm$4\% }  & \multicolumn{1}{|c|}{ $\pm$4\% } &
\multicolumn{1}{|c|}{ $\pm$0.5\% }  & \multicolumn{1}{|c|}{ $\pm$1\% } \\
\multicolumn{1}{|l||}{ Background } & 
\multicolumn{1}{|c|}{ $\pm$57\% }  & \multicolumn{1}{|c|}{ $\pm$0.1\% } &
\multicolumn{1}{|c|}{ $\pm$55\% }  & \multicolumn{1}{|c|}{ $\pm$2\% } &
\multicolumn{1}{|c|}{ $\pm$62\% }  & \multicolumn{1}{|c|}{ $\pm$5\% } \\
\hline
\multicolumn{1}{|l||}{ Total } & 
\multicolumn{1}{|c|}{ }  & \multicolumn{1}{|c|}{ $\pm$8\% } &
\multicolumn{1}{|c|}{ }  & \multicolumn{1}{|c|}{ $\pm$8\% } &
\multicolumn{1}{|c|}{ }  & \multicolumn{1}{|c|}{ $\pm$9\% } \\
\hline
\end{tabular}
\end{center} 
\caption{Contributions to systematic error in the cross-section measurement 
of the process $e^+e^-\rightarrow\nu\bar{\nu}\gamma$.
The total systematic error is the quadratic sum of the individual errors.} 
\label{syst} 
\end{table}

\subsection{Non-pointing single-photon events and multi-photon events} 

The numbers of events with a single non-pointing photon or with
multi-photon final states are compared to Standard Model expectations
in Table~\ref{samples}.

\begin{figure}[hbt]
\begin{center}
\mbox{\epsfxsize 11cm \epsfbox{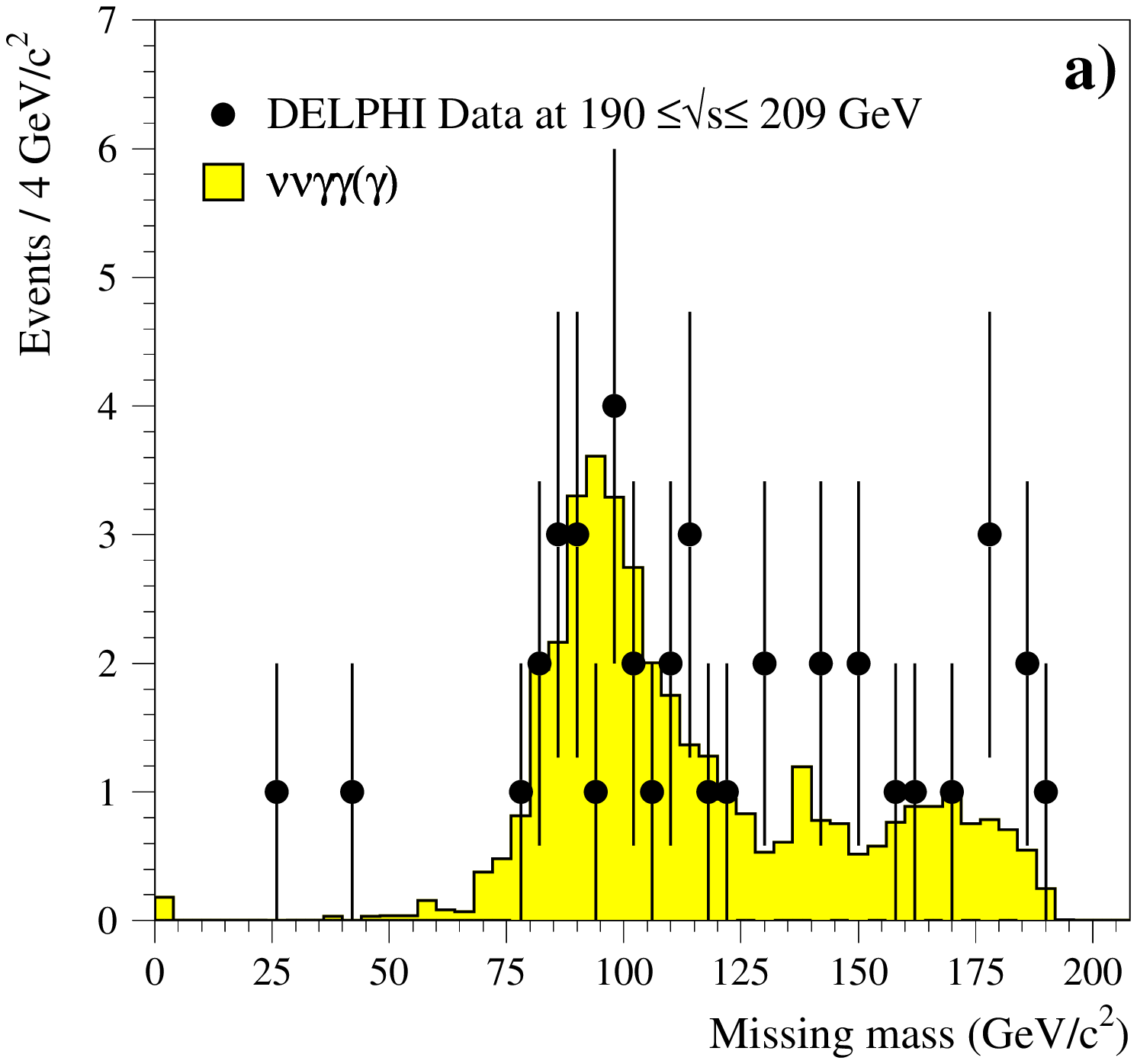}}
\mbox{\epsfxsize 11cm \epsfbox{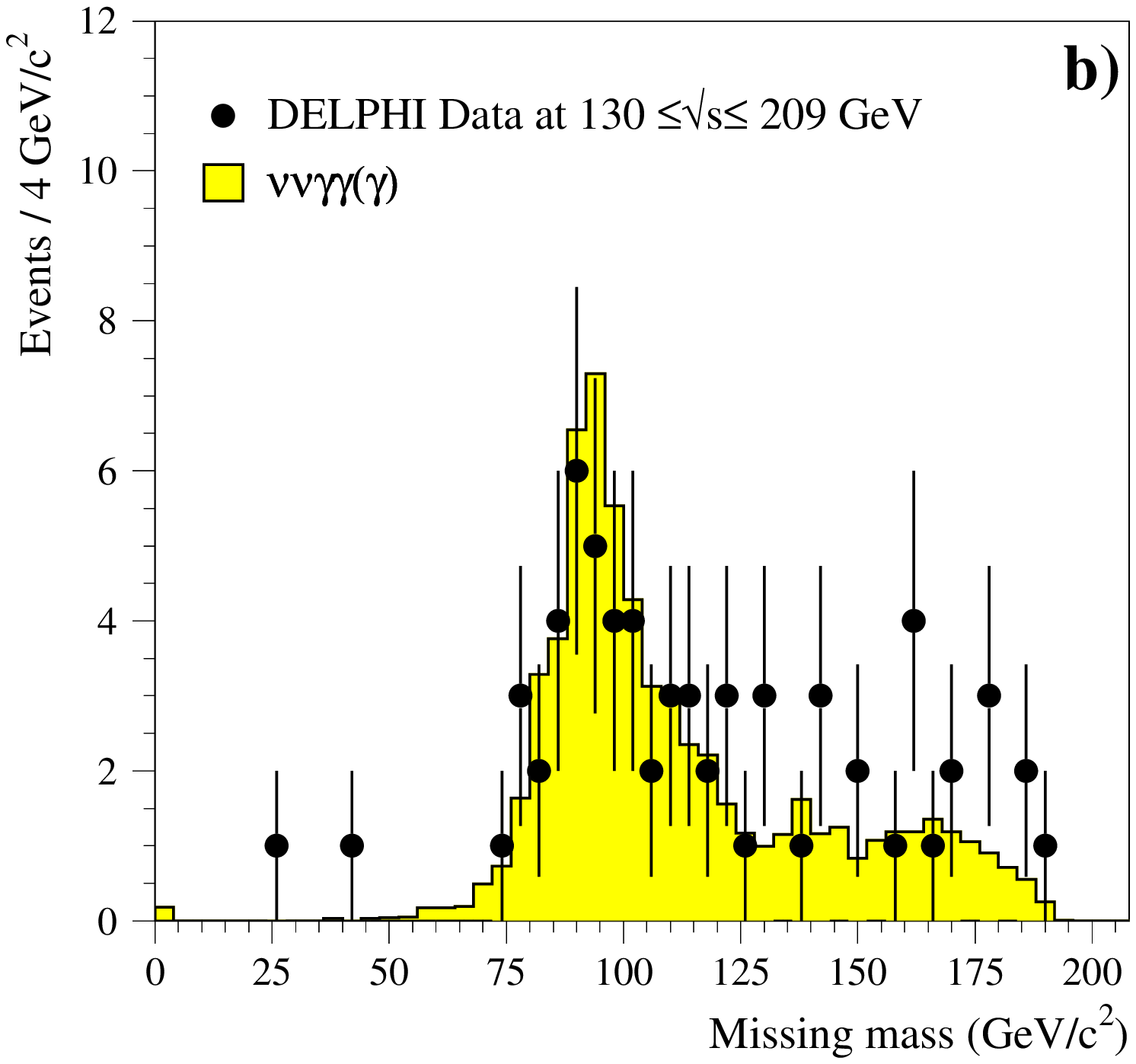}}
\end{center}
\caption[]{Missing mass distribution observed after multi-photon 
preselection in the 190-209~GeV sample
(a) and the combined 130-209~GeV sample (b).}
\label{preselection}
\end{figure}

The missing mass spectra for the preselected multi-photon events and
the expected contribution from
$e^+e^-\rightarrow\nu\overline{\nu}\gamma\gamma(\gamma)$ as simulated
with KORALZ are shown in Figure~\ref{preselection}.  Additional
background contributions from the processes $e^+e^- \rightarrow
e^+e^-\gamma$ and $e^+e^- \rightarrow \gamma \gamma \gamma$ have been
estimated to be 0.43$\pm$0.16 events at the preselection level and
have been added to the simulated sample.  The measured missing mass
distribution is in good agreement with the simulation and no
significant excess over Standard Model expectations was found in any
of the data samples collected at energies up to $\sqrt{s}=209$~GeV.


\begin{table}[t]
\begin{center}
\begin{tabular}{|c|c|c|c|c|}
\cline{2-5}
\multicolumn{1}{c|}{ }    & 
\multicolumn{2}{|c||}{ 130-190~GeV }  & 
\multicolumn{2}{|c|}{ 190-209~GeV }  \\
\cline{2-5}
\multicolumn{1}{c|}{  } & 
\multicolumn{1}{|c|}{ Observed }      & \multicolumn{1}{|c||}{ Expected } &
\multicolumn{1}{|c|}{ Observed }      & \multicolumn{1}{|c|}{ Expected } \\
\hline
\multicolumn{1}{|c|}{ Preselected multi-photon events } & 
\multicolumn{1}{|c|}{ 27 }      & \multicolumn{1}{|c||}{ 25.3$\pm$1.0 } &
\multicolumn{1}{|c|}{ 41 }      & \multicolumn{1}{|c|}{  39.3$\pm$0.8 } \\
\hline
\multicolumn{1}{|c|}{ $\eeGgGg$ selection } & 
\multicolumn{1}{|c|}{ 7 }      & \multicolumn{1}{|c||}{ 7.1$\pm$0.5 } &
\multicolumn{1}{|c|}{ 17 }     & \multicolumn{1}{|c|}{ 13.8$\pm$0.4 } \\
\hline
\multicolumn{1}{|c|}{ $\eeXgXg$ selection } & 
\multicolumn{1}{|c|}{ 12 }      & \multicolumn{1}{|c||}{ 8.6$\pm$0.6 } &
\multicolumn{1}{|c|}{ 14 }     & \multicolumn{1}{|c|}{ 15.2$\pm$0.6 } \\
\hline
\multicolumn{1}{|c|}{ Non-pointing single-photon events } & 
\multicolumn{1}{|c|}{ 6 }      & \multicolumn{1}{|c||}{ 7.6$\pm$0.9 } &
\multicolumn{1}{|c|}{ 10 }     & \multicolumn{1}{|c|}{  7.0$\pm$0.7 } \\
\hline
\end{tabular}  
\end{center} 
\caption[]{The number of observed and expected events from Standard Model sources 
in four selected data samples.}
\label{samples} 
\end{table}

\section{Limits on new phenomena}

\subsection{Neutrino physics}

\subsubsection{Measurement of the number of light neutrino families}

A measurement of the cross-section of the process $e^+e^- \rightarrow
\nu \bar{\nu}\gamma$ determines the number of light neutrino
generations, $N_{\nu}$.  The LEP2 cross-section measurements have been
compared with the expected cross-sections for 2, 3 and 4 neutrino
generations, calculated with KORALZ, and the number of neutrino
generations has been deduced (Table~\ref{cross}).  Averaging the
independent measurements from the three different calorimeters at
$\sqrt{s}=180-209$ GeV, the number of light neutrino generations
becomes:  $$N_{\nu}=2.84\pm0.10(stat)\pm0.14(syst)$$

\subsubsection{Neutrino interactions beyond the Standard Model}

The contribution to the single-photon process $\eenng$
from hypothetical Non-Standard (NS) neutrino interactions with the electron, has
recently been calculated in order to set limits on theories which attempt
to explain the solar and atmospheric neutrino anomalies~\cite{nsn}.  
The single-photon cross-section due to these non-standard neutrino interactions
was computed at tree level in a point interaction approximation where
$\epsilon_{\alpha R}$ and $\epsilon_{\alpha L}$ ($\alpha = e, \mu,
\tau$) parameterise the strength, relative to the Fermi
constant ($G_{F}$), of the non-standard interactions
between the three neutrinos and the electron.  
The radiator approximation~\cite{ISR} was used in these
calculations to determine the single-photon cross-section from initial
state radiation.

The analysis was made for three different assumptions as suggested
by~\cite{nsn}:

\begin{enumerate}
\item The NS interactions couple only to the electron neutrino, 
$e^+e^-\rightarrow \nu_{e}\bar{\nu}_{e}\gamma$.
\item The NS interactions couple only to the tau neutrino, 
$e^+e^-\rightarrow \nu_{\tau}\bar{\nu}_{\tau}\gamma$.
\item The NS interactions couple only to a flavour changing 
neutral current, 
$e^+e^-\rightarrow \nu_{\alpha}\bar{\nu}_{\beta}\gamma$ where $\alpha \neq \beta$.
\end{enumerate}

The analysis was in all three cases based on the single-photon
samples recorded with the HPC and FEMC detectors and the event
selection was optimised with the likelihood ratio method mentioned 
in section 3.1.2.  Under the
first assumption, the interference terms between the non-standard interaction
and the Standard Model amplitudes can, in some parts of the $\epsilon_{e L} -
\epsilon_{e R}$ parameter space, give a negative contribution to the
single-photon cross-section at a level which the analysis is sensitive
enough to exclude.  
The analysis under the first
assumption was for this reason divided into two independent parts based on the
regions of the photon energy distribution for which a positive and a negative
contribution from non-standard neutrino interactions was predicted.

\begin{figure}[hbt]
\begin{center}
\mbox{\epsfxsize 7.9cm \epsfbox{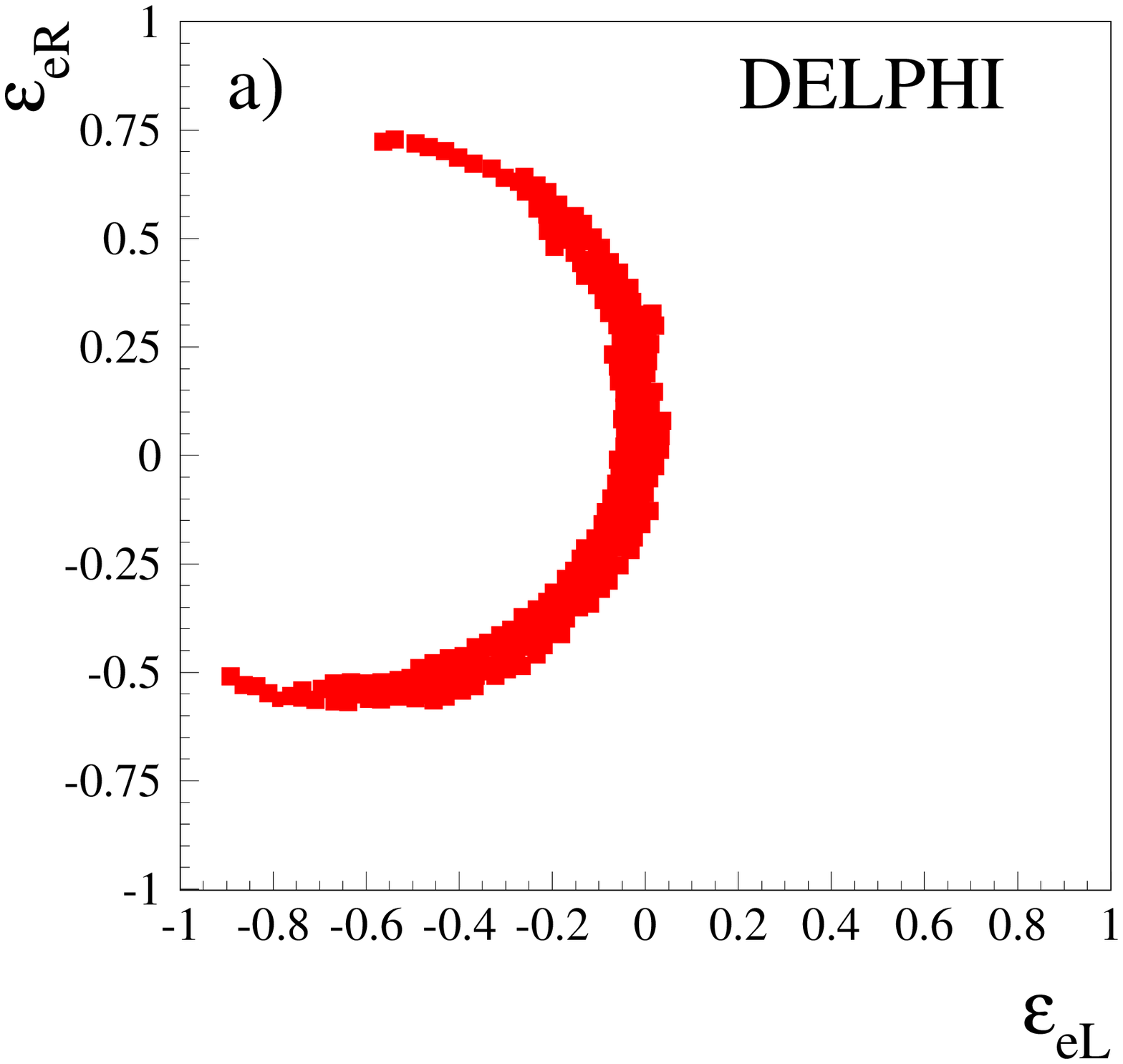}}
\mbox{\epsfxsize 7.9cm \epsfbox{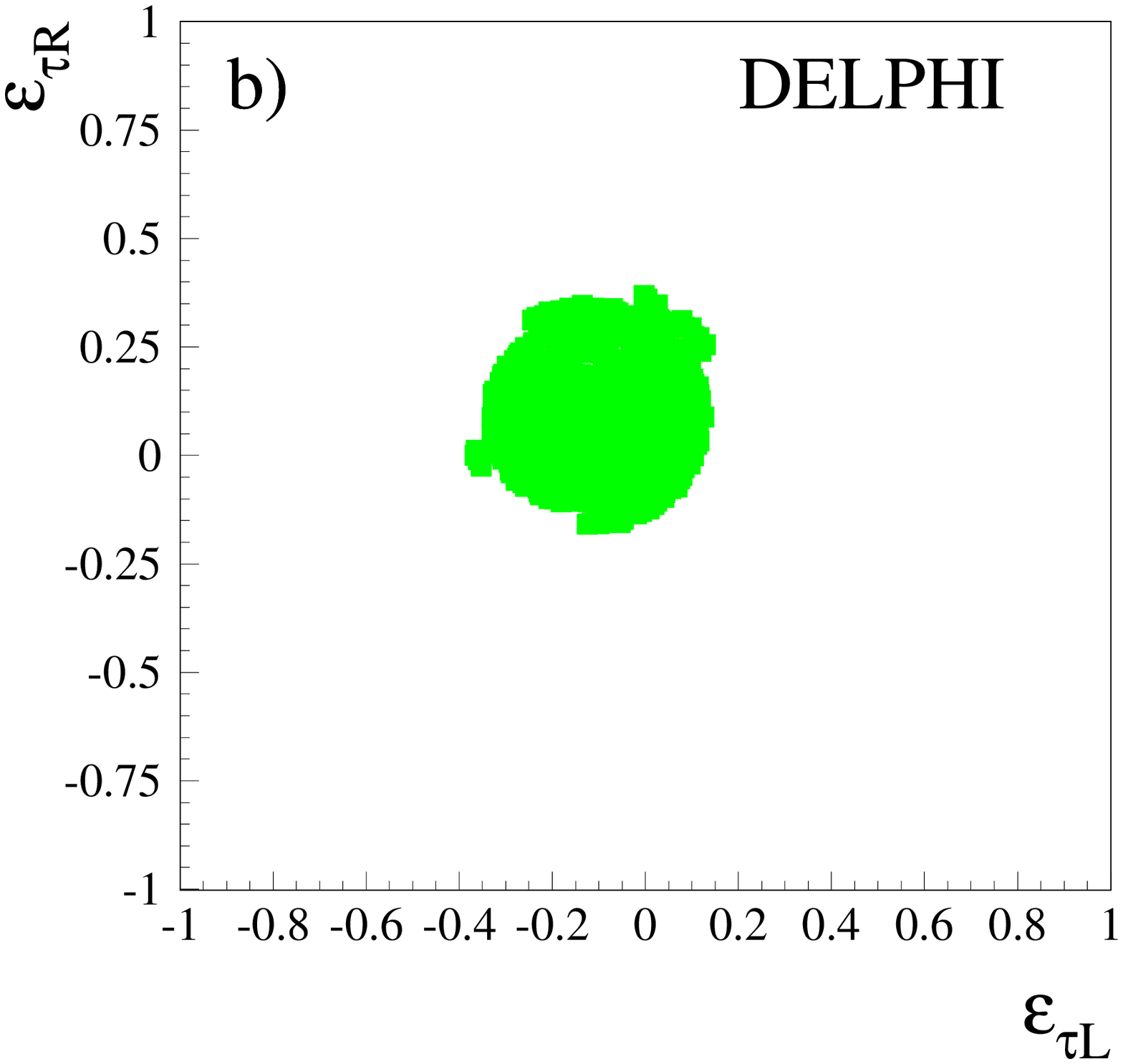}}
\mbox{\epsfxsize 7.9cm \epsfbox{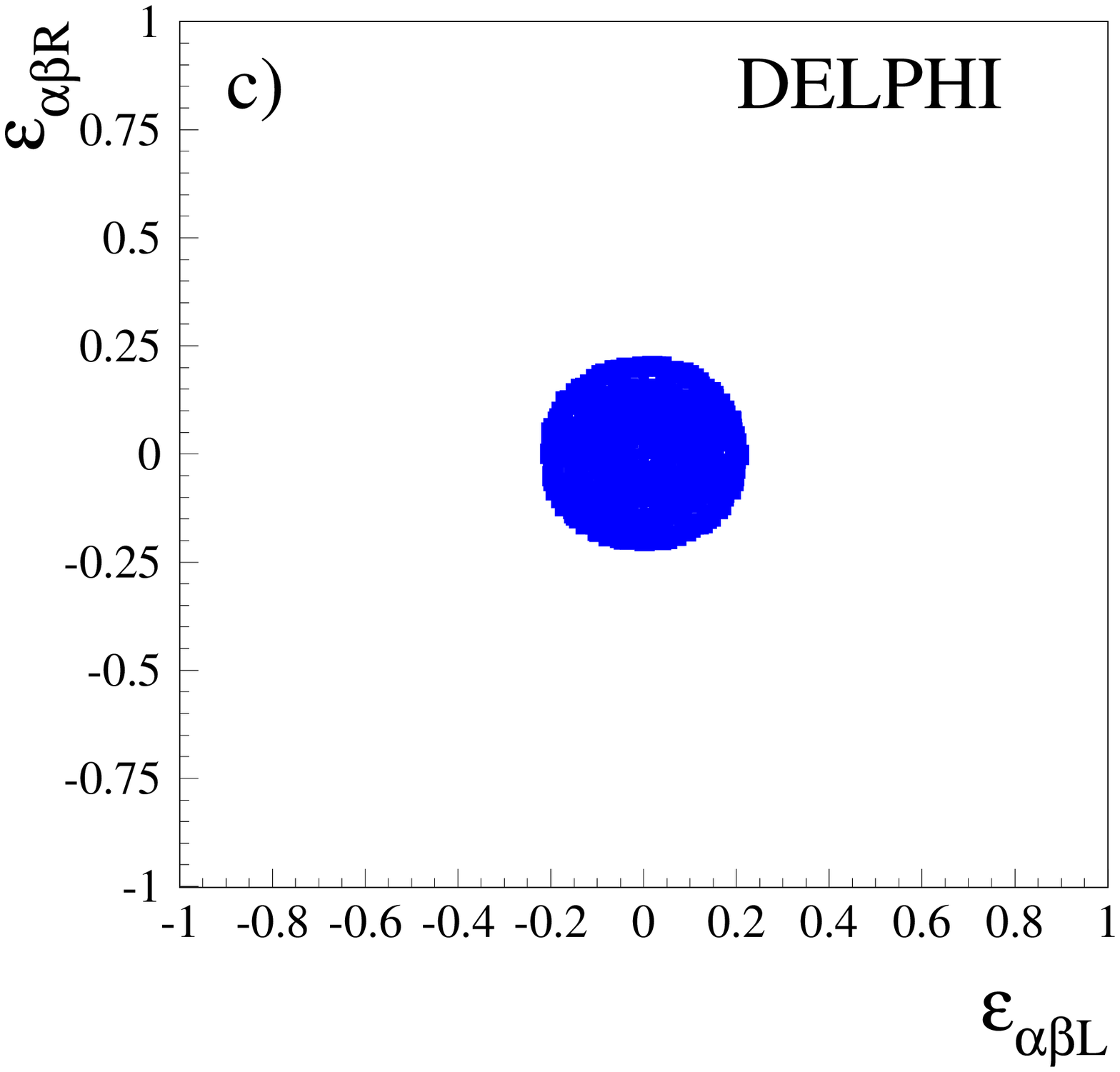}}
\end{center}
\caption[]{
The remaining allowed regions. 
a) In the $\epsilon_{e L} - \epsilon_{e R}$ plane.
b) In the $\epsilon_{\tau L} - \epsilon_{\tau R}$ plane.
c) In the $\epsilon_{\alpha\beta L} - \epsilon_{\alpha\beta R}$ ($\alpha \neq \beta$) 
plane.
} 
\label{eppntfc} 
\end{figure}

The calculated allowed and excluded regions in the $\epsilon_{L} - \epsilon_{R}$ 
planes for
the three different assumptions are shown in Figure~\ref{eppntfc}.  To
exclude a point in the $\epsilon_{e L} - \epsilon_{e R}$ plane a
logical OR was used between the two analyses of positive and negative
interference and this gives the ``crescent moon'' shape of the
allowed region in Figure~\ref{eppntfc}a.  All limits were calculated at
a 95\% C.L. and the limits were computed by
combining all the FEMC and HPC data samples in Table~\ref{lumi}.

\subsection{Limits on the production of unknown neutral states}

\subsubsection{Limits on $e^+e^-\rightarrow X\gamma$ production}

\begin{figure}[hbt]
\begin{center}
\mbox{\epsfxsize 11cm \epsfbox{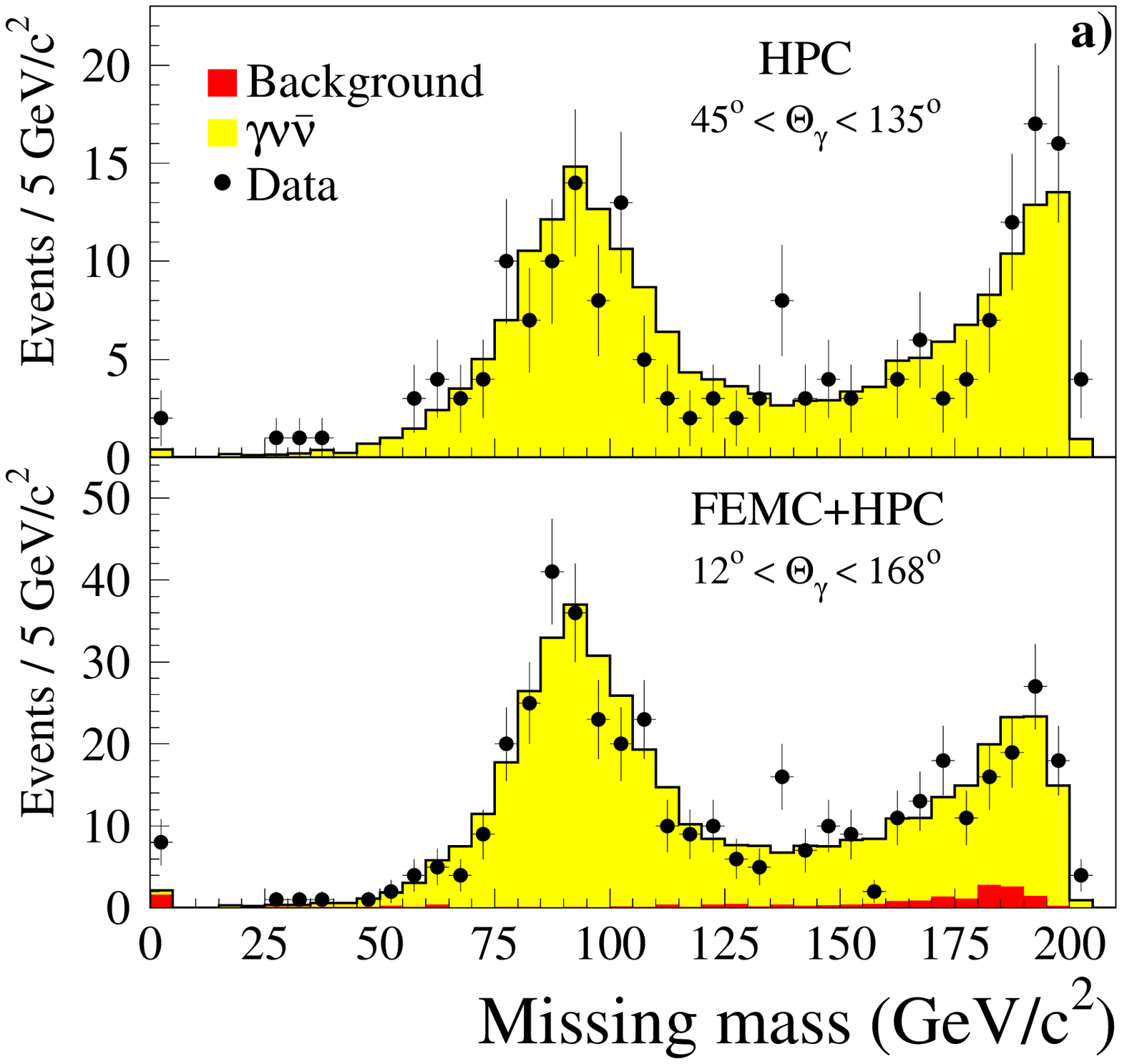}}
\mbox{\epsfxsize 11cm \epsfbox{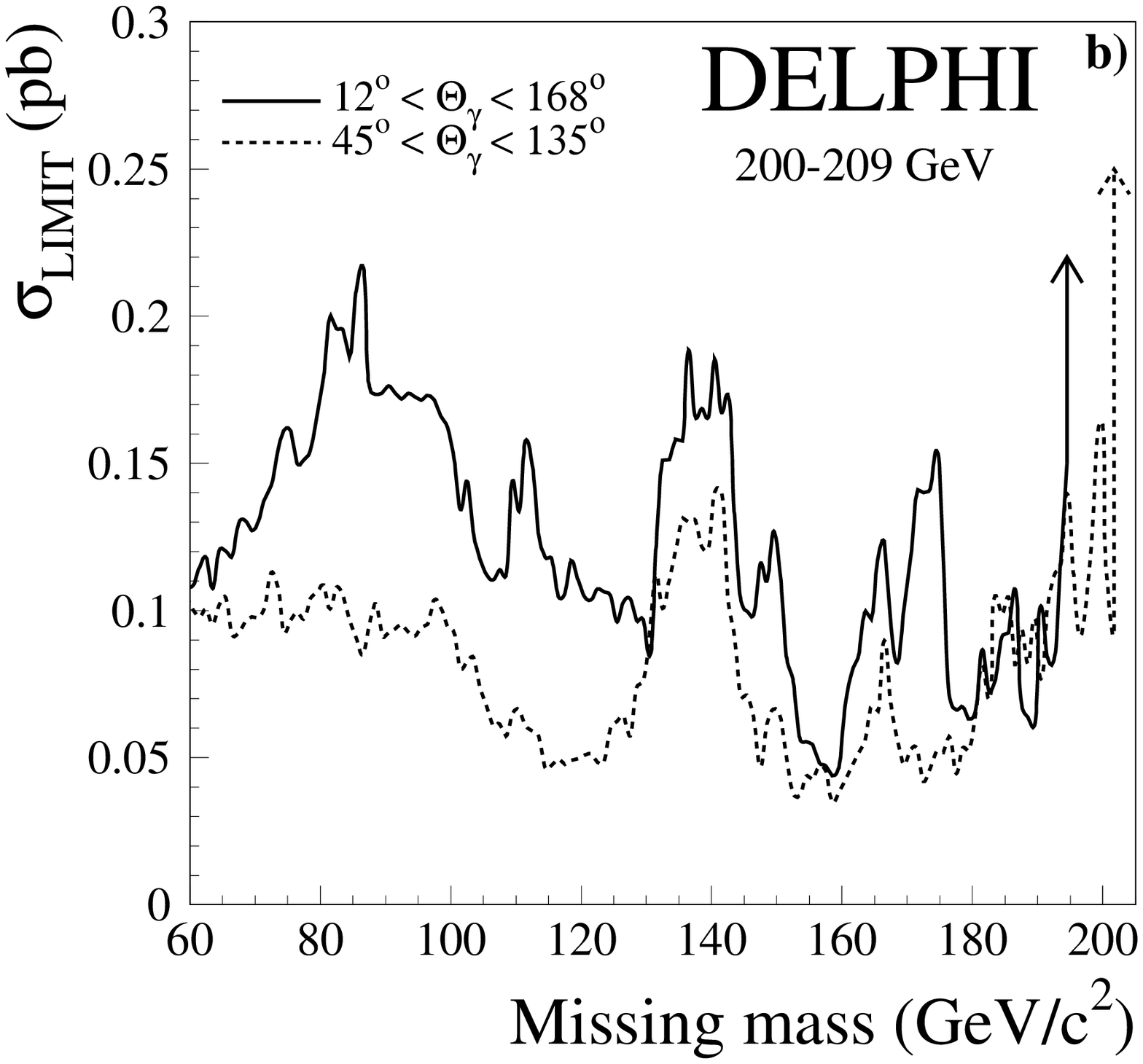}}
\end{center}
\caption[]{a) The distributions of the missing mass 
for the events at 200-209~GeV in the HPC and in the FEMC+HPC.
The light shaded area is the expected distribution from
$e^+e^- \rightarrow \nu \bar{\nu}\gamma$ and the
dark shaded area is the total background from other
sources. 
b) Upper limits at 95$\%$~C.L. and at 205~GeV
(within the quoted solid angles) 
for the production of a new unknown stable neutral object .
}
\label{mrec}
\end{figure}

The observed single-photon events have been used to set a limit on the
production cross-section of a new hypothetical particle, X, produced
in association with an ISR photon and being stable or decaying to
invisible decay products. The upper limits for the cross-section 
of the
process $e^+e^- \rightarrow \gamma + X$ were 
calculated from the missing mass
distribution (Figure~\ref{mrec}) at $\sqrt{s}=200-209$~GeV (average
$\sqrt{s}=$205.4~GeV) of 
the 190 events in the angular region $45^\circ-135^\circ$ and
the 414 single $\gamma$ events in the angular
region $12^\circ-168^\circ$ while taking into account the expected
background.  The limits are valid if the intrinsic width of the $X$
particle is negligible compared to the detector resolution (the recoil
mass resolution varies between 10 GeV/c$^2$ at the $Z^0$ peak to 1 GeV/c$^2$ at
high masses). Separate 95\% C.L. upper limits are given in Figure~\ref{mrec} 
for photons in the HPC region and in the FEMC+HPC
region.  In the latter case an assumption of an ISR-like photon
angular distribution has been made to correct for losses between the
calorimeters.

\subsubsection{Limits on $e^+e^-\rightarrow XY \rightarrow YY\gamma$ production}

A cross-section limit has been calculated for the scenario where
hypothetical $X$- and $Y$-particles are produced and the $X$-particle
decays into a $Y$ and a photon, $e^+e^- \rightarrow XY \rightarrow
YY\gamma$.  It is assumed that the $Y$ particles escape detection
(such as in the process $e^+e^- \rightarrow
\tilde{\chi}_2^0\tilde{\chi}_1^0 \rightarrow
\tilde{\chi}_1^0\tilde{\chi}_1^0\gamma$ which is predicted by certain
SUSY models) and that the branching ratio of $X \rightarrow Y\gamma$
is 100\% .

\begin{figure}[hbt]
\begin{center}
\mbox{\epsfxsize 11cm \epsfbox{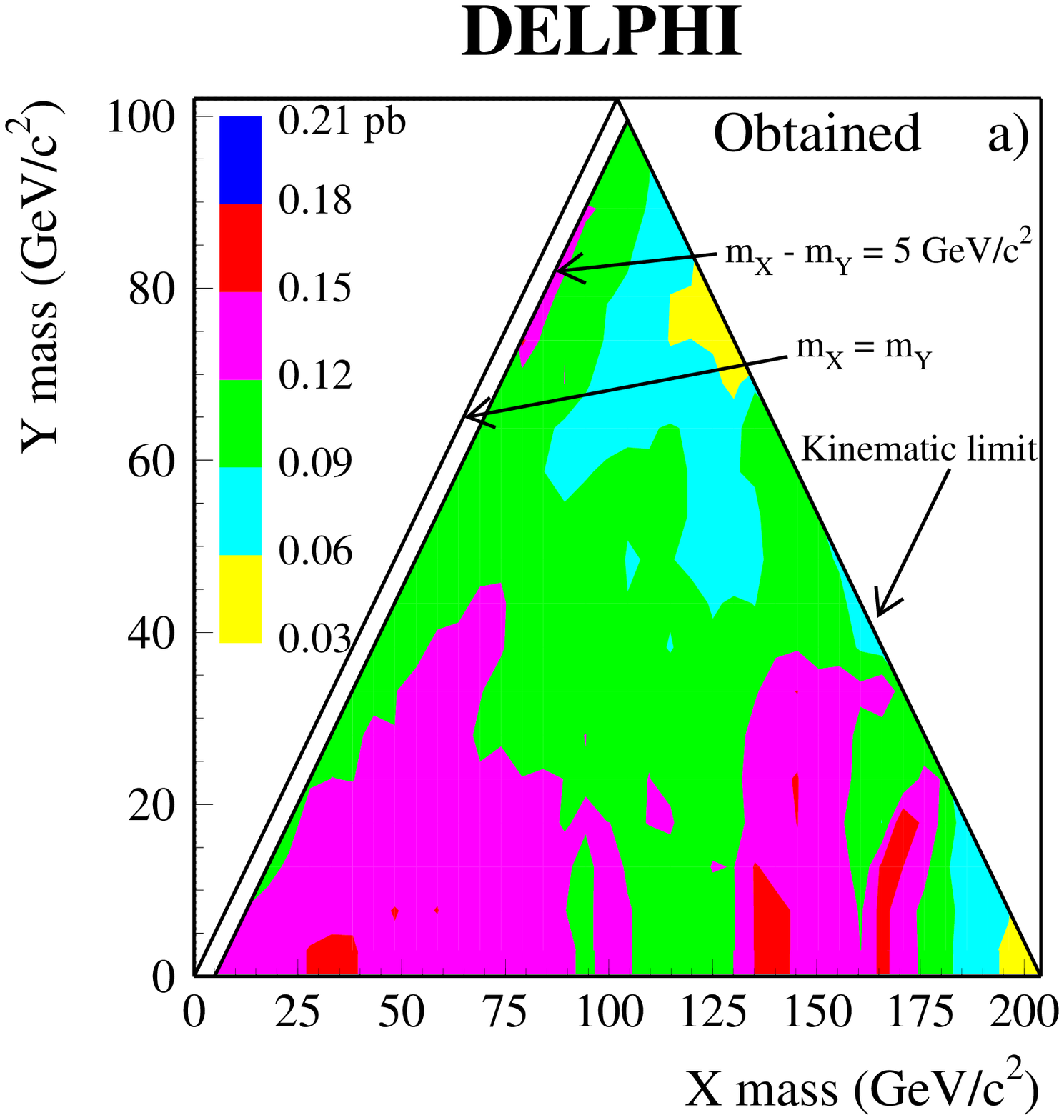}}
\mbox{\epsfxsize 11cm \epsfbox{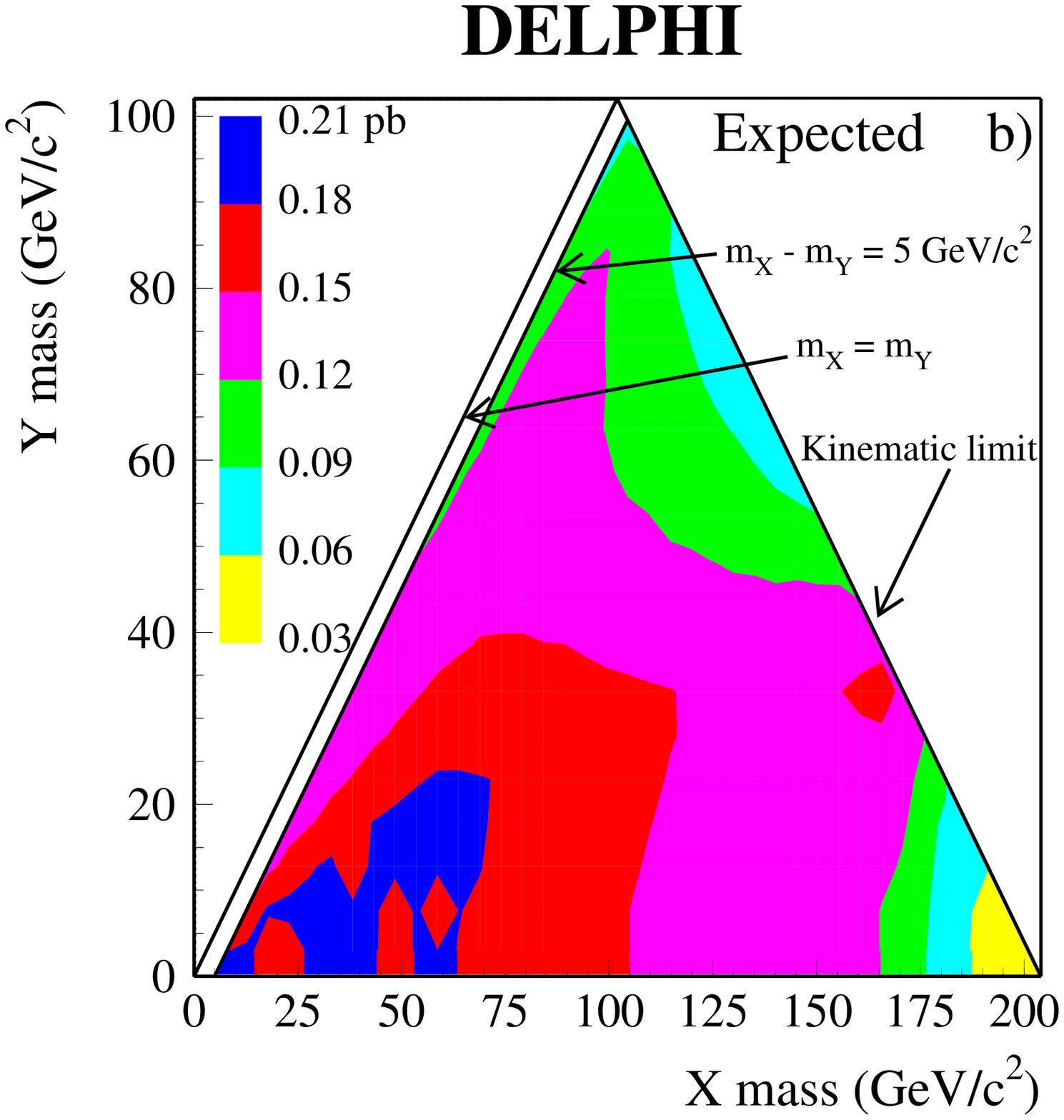}}
\end{center}
\caption[]{The obtained (a) and expected (b) cross-section limit at 95$\%$~C.L. 
and at 204 GeV for the process $e^+e^-\rightarrow XY \rightarrow YY\gamma$ 
where $X$ and $Y$ are hypothetical new neutral particles.}
\label{xy}
\end{figure}

It is, in addition, assumed that the $E_{\gamma}$ and
$\cos\theta_{\gamma}$ distributions are flat.  This assumption
results in a very small predicted signal within the STIC acceptance
and for this reason only the single-photon events recorded by the HPC
and FEMC detectors were used in this analysis.  The data collected
at $\sqrt{s}$=199-209 GeV (average$\sqrt{s}=$204.0~GeV)
were used in the analysis and it was
assumed that any $\sqrt{s}$-dependence of the signal would be
insignificant in this limited $\sqrt{s}$-range.  The analysis has been
performed at 401 points in the mass plane of the $X$ and $Y$ particles
where the selection cuts have been optimised at every point with a
likelihood ratio method.  The obtained and expected cross-section
limits, within the HPC + FEMC acceptance, in the mass plane of the $X$
and $Y$ particles are shown in Figure~\ref{xy}.

\subsection{Limits on the production of gravitons}

If there are extra compact dimensions of space in which only gravity
can propagate, gravitational interactions could be unified with gauge
interactions already at the weak scale~\cite{grav1,grav2}. The
consequence of this model is that at LEP gravity could manifest itself
by the production of gravitons ($G$), which themselves would be
undetectable by the experiments.  Instead single photons from the
$e^+e^-\rightarrow\gamma G$ reaction could be observable.

In these gravitational models, a fundamental mass scale, $M_{D}$, 
is introduced, which is related to the gravitational constant 
$G_{N}$ and to the size or radius $R$ of the compactified space
(assumed to be a torus) by
$ M_{D}^{n+2}R^n = (4\pi G_N)^{-1}$~\cite{grav1} or
$ M_{D}^{n+2}R^n = (8\pi G_N)^{-1}$~\cite{grav2}
where $n$ is the number of dimensions in addition to the usual
3+1-dimensional space\footnotemark . 

\footnotetext{Note that the definition of M$_D$ in~\cite{grav1} is a factor $2^{1/(n+2)}$ 
larger than in~\cite{grav2} and that the limits presented here are for an 
M$_D$ as defined by~\cite{grav2}. }

With one extra dimension and a fundamental
scale of 0.5-1~TeV, the size of this dimension becomes
$10^{12}-10^{13}$~m which is excluded by macroscopic measurements.
However, already with two extra dimensions, $R$ is in the range 0.5-2~mm
and with $n$=6 the size of the dimensions becomes 30-70~fm. 
In this case the modification of the gravitational force
would not have been observed in previous gravitational measurements.
 
The differential
cross-section for the $e^+e^-\rightarrow\gamma G$ process has been
calculated in~\cite{grav2}. Initial state radiation can produce additional
photons that would cause a signal event to be rejected in a single-photon 
analysis. The expected signal cross-section has therefore been corrected
with a radiator approximation method~\cite{ISR}. 

\begin{figure}[hbt]
\centerline{\epsfig {file=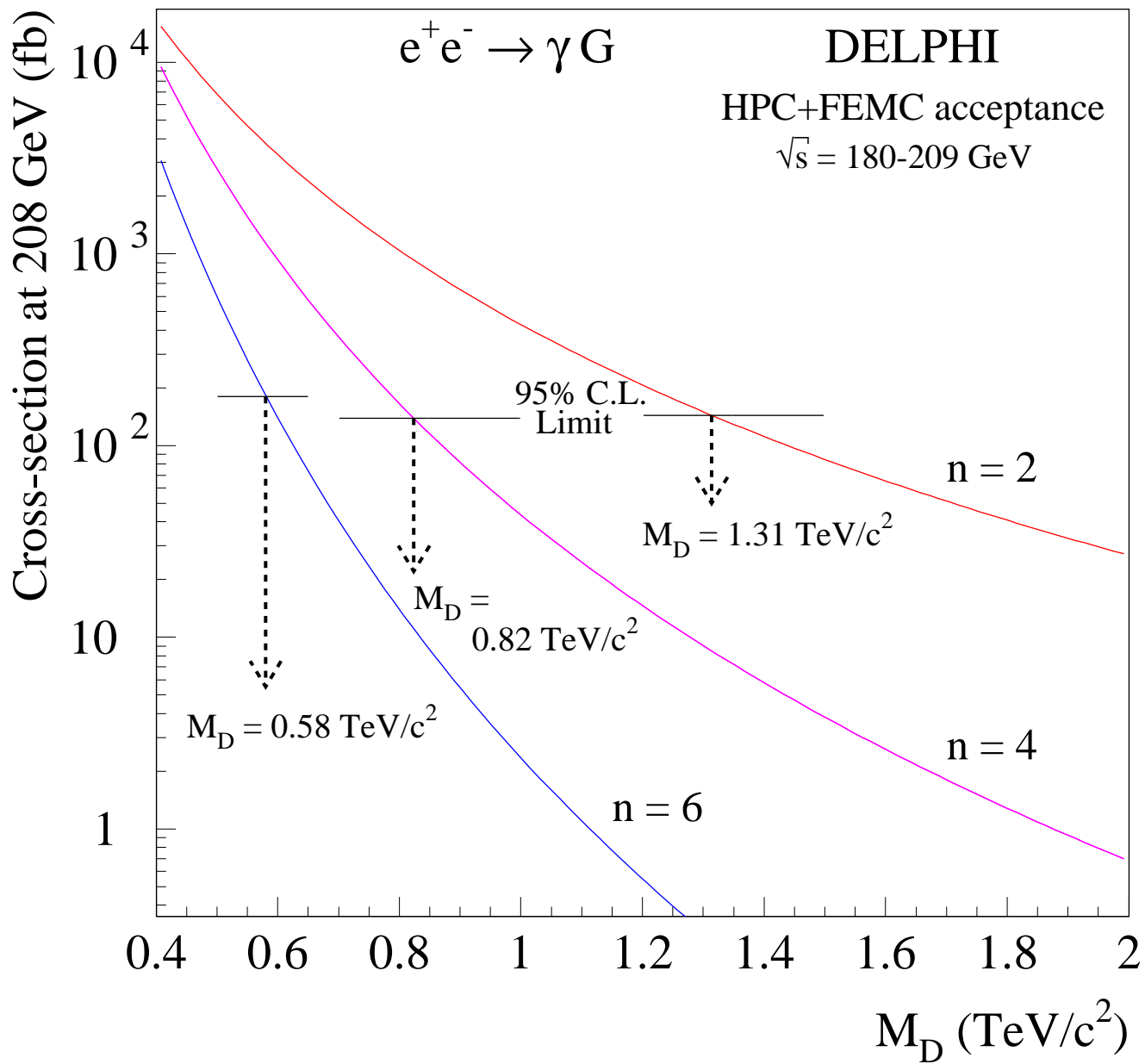,width=18cm}}
\caption[]{The cross-section limit at 95\% C.L. 
for $e^+e^-\rightarrow\gamma G$ production at \linebreak $\sqrt{s}$=208~GeV
and the expected cross-section for 2, 4 and 6 extra dimensions.} 
\label{grav} 
\end{figure}

From Figure~\ref{missing_mass} it is clear that only a very small signal
can be expected in the STIC detector compared with the one in the FEMC
and the HPC.  For this reason only the latter two detectors were used
in this analysis.  All DELPHI data with $\sqrt{s}>180$~GeV were used
and for each bin in $\sqrt{s}$ (see Table~\ref{lumi}) a limit was
calculated after a cut optimisation based on a likelihood ratio
method as described in section 3.1.2.  These limits were then combined to give a
95\% C.L. cross-section limit at 208 GeV of 0.14~pb for 2 and
4 extra dimensions.  The resulting limits on the fundamental mass
scale are $M_{D}>1.31$~TeV/c$^2$ for $n$=2 and $M_{D}>0.82$~TeV/c$^2$ for $n$=4
(Figure~\ref{grav}).  This in turn can be transformed into a limit on the size
(radius) of the extra dimensions which is $R<0.27$~mm and $R<13$~pm for
2 and 4 extra dimensions respectively.  Limits for other numbers of extra
dimensions are given in Table~\ref{extra}. The systematic error on 
the M$_D$ limits is estimated to be between 1\% and 2\% .


\begin{table}[h]
\begin {center}
\begin{tabular}{|c|c|c|c|c|} \hline
Dimensions & $\sigma_{Limit}$ (pb) & M$_D$ (expected) & M$_D$ (obtained) & Radius  \\ \hline
2 & 0.14 & $>$1.27 TeV/c$^2$ & $>$1.31 TeV/c$^2$ & $<$0.27 mm\\
3 & 0.14 & $>$0.98 TeV/c$^2$ & $>$1.02 TeV/c$^2$ & $<$3.5 nm \\
4 & 0.14 & $>$0.80 TeV/c$^2$ & $>$0.82 TeV/c$^2$ & $<$13 pm \\
5 & 0.17 & $>$0.67 TeV/c$^2$ & $>$0.67 TeV/c$^2$ & $<$0.49 pm \\
6 & 0.18 & $>$0.58 TeV/c$^2$ & $>$0.58 TeV/c$^2$ & $<$55 fm \\ \hline
\end{tabular}
\end {center}
\caption[]{Limits at the 95\% confidence level on the fundamental 
mass scale M$_D$ and the radius for different numbers of extra dimensions.
}
\label{extra} 
\end{table}

\subsection{SUSY particles}

\subsubsection{Limits on the gravitino mass}

The cross-section for the process $\eeGG$ has been computed under the
assumption that all other supersymmetric particles are too heavy to be
produced~\cite{gravitino}. The largest sensitivity in this type of search 
is obtained with photons at low energy and low polar angle. Since the
signal cross-section grows as the sixth power of the centre-of-mass
energy, the highest sensitivity is also obtained at the highest available
beam energy. Lower limits on the mass of a gravitino 
produced by the $\eeGG$ process have been extracted previously at
LEP~\cite{delphi_189,LEP_results}. 

The expected signal cross-section in~\cite{gravitino} 
has in the present study been corrected for
initial state radiation~\cite{ISR} that produces multi-photon events
which are rejected in the analysis. The large
energy cut in the STIC analysis meant that this detector did not contribute
significantly to the results.
Limits were therefore calculated for only the FEMC and HPC data in the ten
$\sqrt{s}$ bins of Table~\ref{lumi} and after a similar combination of these limits
as in the graviton analysis, a 95\% C.L. 
limit of $\sigma<0.18~\rm{pb}$ at 208 GeV was
obtained.  This corresponds to a lower limit on the gravitino
mass which is $$m_{\tilde{{G}}}> 1.09 \cdot 10^{-5} ~
\rm{eV/c^2}{\mbox{~~~at 95\% C.L.}}$$ while the expected limit is
$1.10 \cdot 10^{-5} ~ \rm{eV/c^2}$. The systematic error on the
$m_{\tilde{{G}}}$ limit is estimated to be 2\% .
Since the supersymmetry-breaking
scale $|F|^{\frac{1}{2}}$ is related to the gravitino mass and the
gravitational constant ($G_N$) by $|F|=\sqrt{\frac{3}{8\pi}/G_N}\cdot
m_{\tilde{{G}}}$, the limit on this scale is
$|F|^{\frac{1}{2}}>214$~GeV.

\subsubsection{Limits on neutralino production if the $\tilde{G}$ is the LSP}

Supersymmetric models such as the gauge-mediated supersymmetric 
model~\cite{GMSB} or the no-scale supergravity model~\cite{LNZ} (LNZ) predict 
that the gravitino $\tilde{G}$ is
the lightest supersymmetric particle (LSP).  If the next-to-lightest
supersymmetric particle (NLSP) is the neutralino $\tilde{\chi}^0_1$,
both single-photon and multi-photon production can occur at LEP2 via
the processes $\eeGGg$ and $\eeGgGg$.  While the rate of the former
process is proportional to the inverse of the gravitino mass squared,
the two-photon process is independent of the gravitino mass.
Consequently, the single-photon process is expected to dominate only
for ultra light gravitinos.

The expected photon distributions from the process $\eeGGg$ were
generated with SUSYGEN~\cite{susygen}, and the event selection was
optimised with a likelihood ratio method.  The same analysis was
repeated at 26 different neutralino masses ($m_{\tilde{\chi}^0_1}$)
between 80 and 208 GeV/c$^2$. The 95\% C.L. cross-section limit for $\eeGGg$
production was computed at 208 GeV after combining
the limits from the single-photon data recorded with the FEMC and
HPC detectors at a $\sqrt{s}$ between 180 and 209 GeV, assuming the
signal cross-section to scale as $\beta^8$ (where $\beta$ is the
neutralino velocity).

\begin{figure}[hbt]
\begin{center}
\mbox{\epsfxsize 12cm \epsfbox{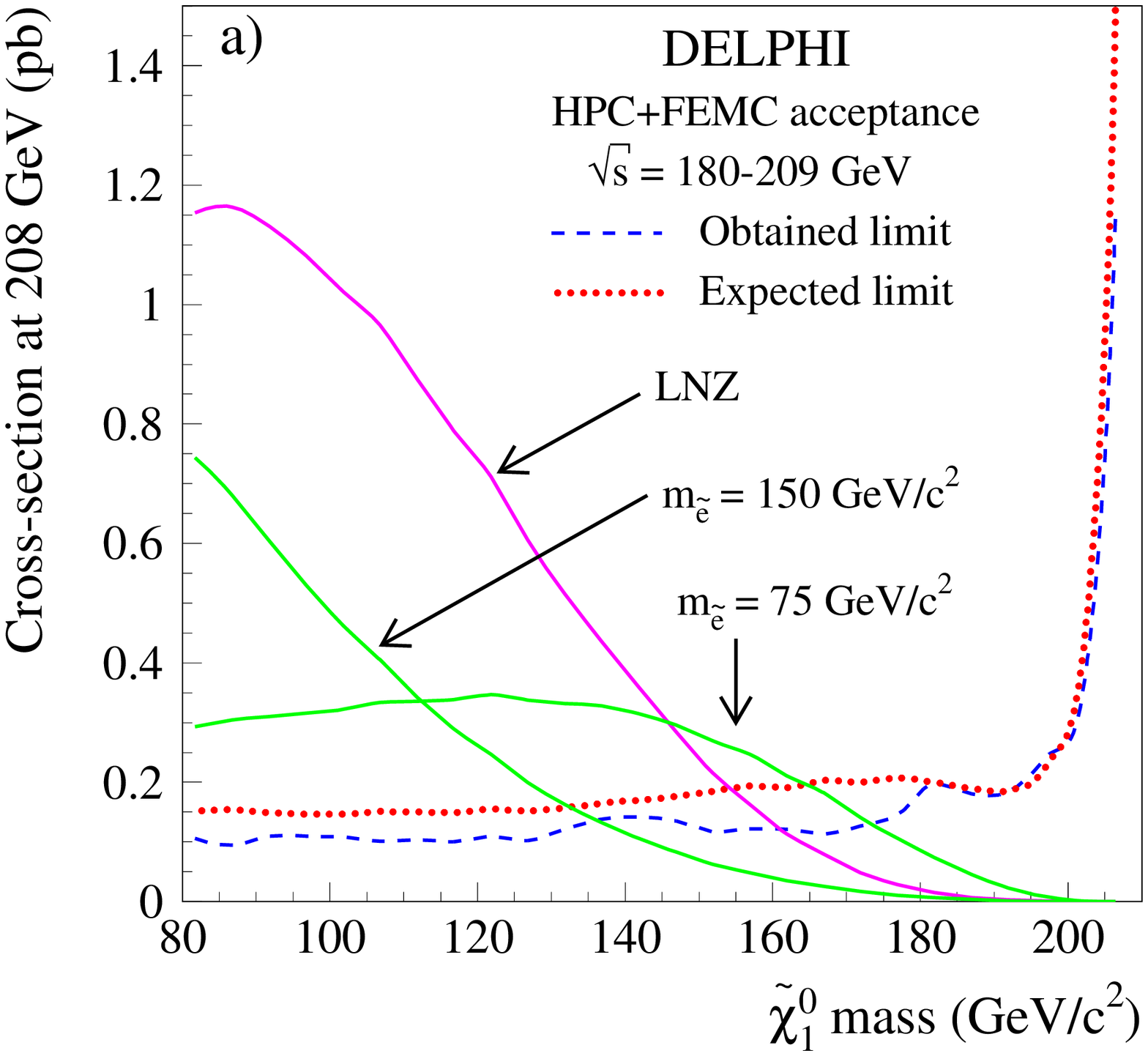}}
\mbox{\epsfxsize 7.8cm \epsfbox{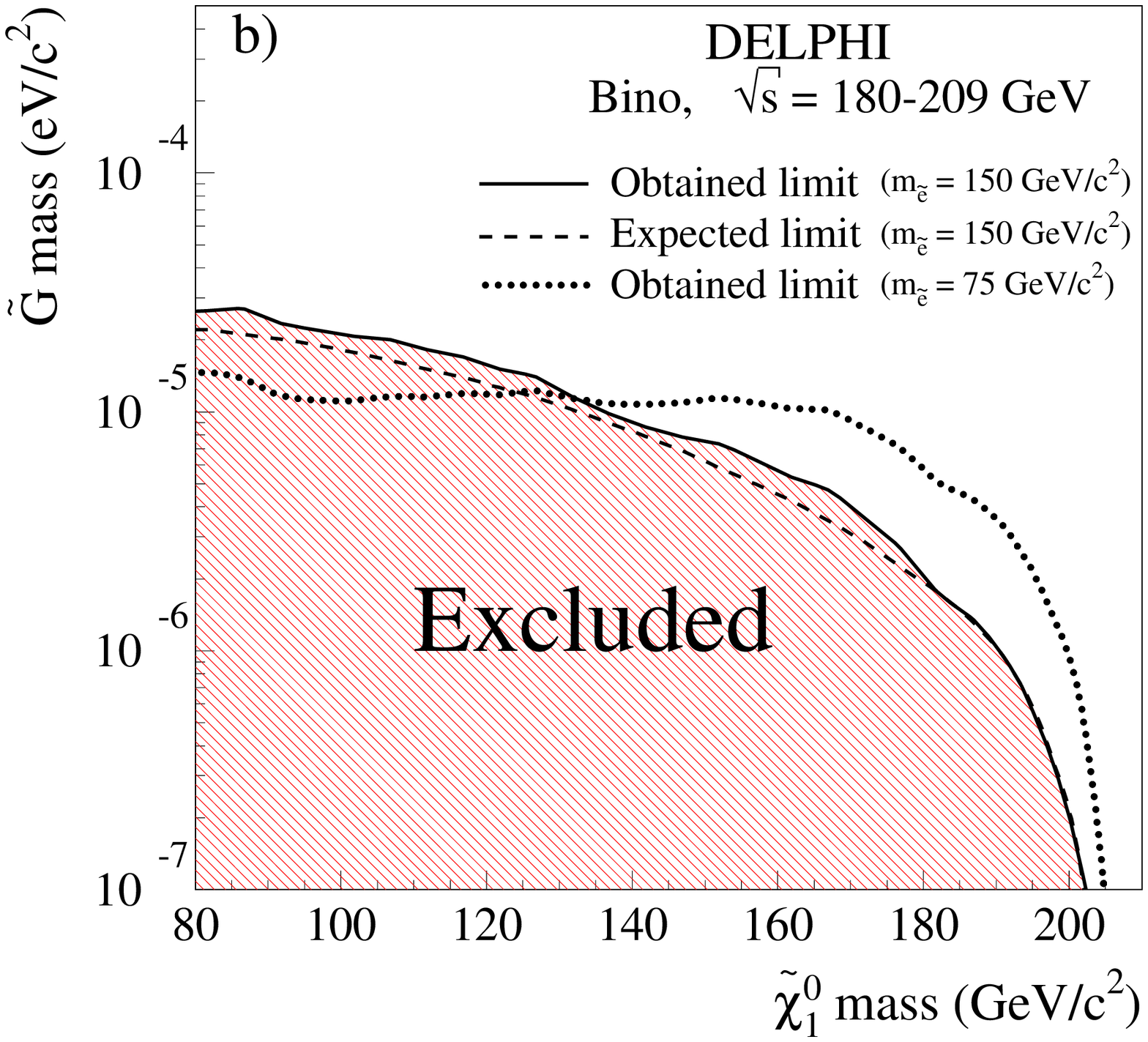}}
\mbox{\epsfxsize 7.8cm \epsfbox{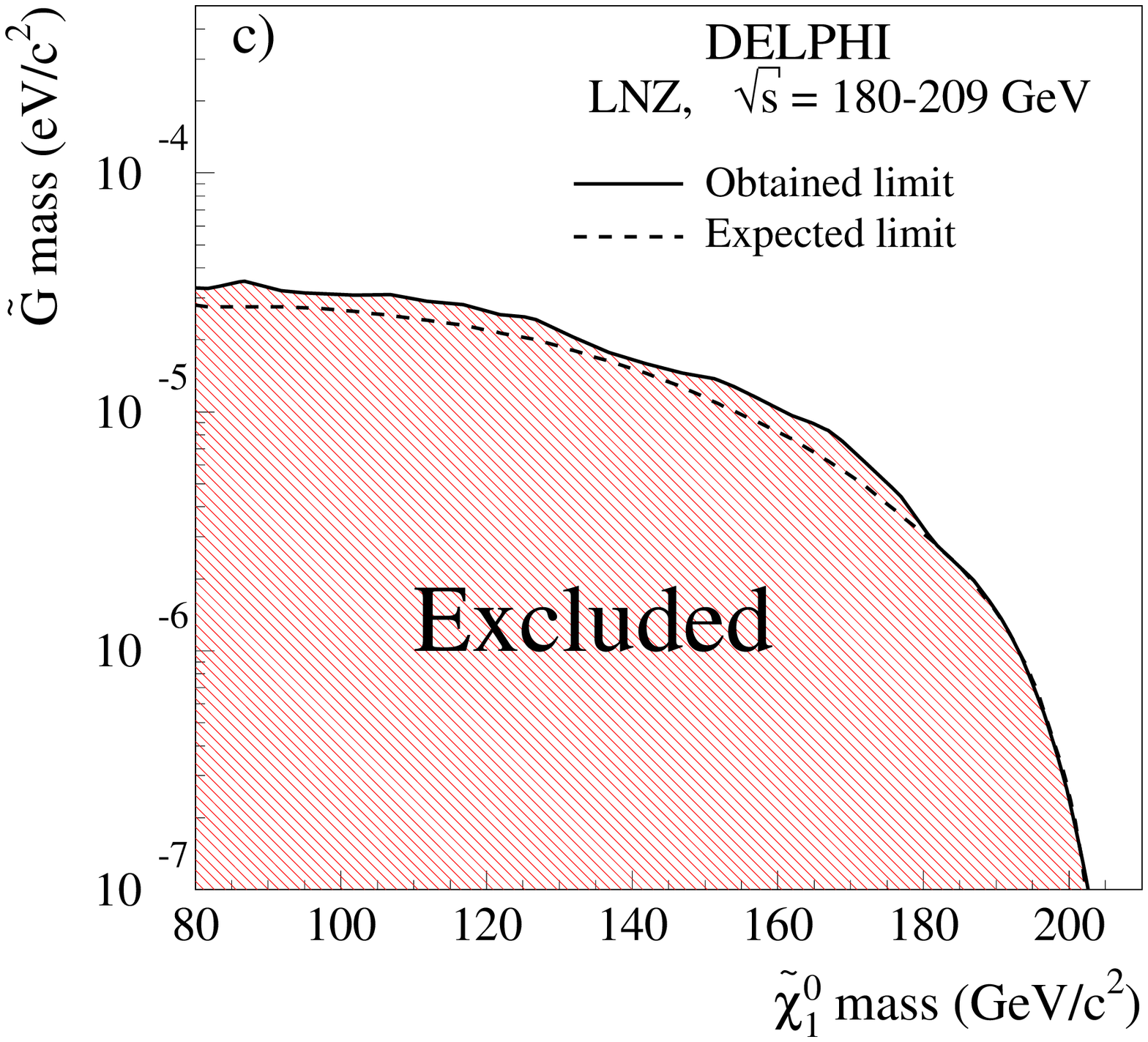}}
\end{center}
\caption[]{a) Upper limit at 95\%~C.L. on the cross-section at $\sqrt{s}=$208~GeV of the process $\eeGGg$
as a function of the $\tilde{\chi}^0_1$ mass. The predicted cross-sections under the assumption
that the neutralino is a Bino or as described by the LNZ-model are also shown for 
$m_{\tilde{G}} = 1\times10^{-5}$ eV/c$^2$.
b), c) Exclusion plots in the $m_{\tilde{\chi^0_1}}$-$m_{\tilde{G}}$ mass plane. 
} 
\label{nlz} 
\end{figure}

The analysis was applied to two different theoretical scenarios. In the
first one, the neutralino was assumed
to be a pure bino and the right- and left-handed selectrons were
degenerate in mass.  Figure~\ref{nlz}a shows the cross-section limit
within the FEMC + HPC acceptance assuming that the branching ratio
$Br(\tilde{\chi}^0_1 \rightarrow \tilde{G}\gamma) =$ 100\% .  The
exclusion regions in the $m_{\tilde{\chi^0_1}}$-$m_{\tilde{G}}$ mass
plane are depicted in Figure~\ref{nlz}b for the selectron masses
$m_{\tilde{e}} =$ 75 GeV/c$^2$ and $m_{\tilde{e}} =$ 150 GeV/c$^2$.
The branching ratios for 
$\tilde{\chi}^0_1 \rightarrow Z \gamma$ and
$\tilde{\chi}^0_1 \rightarrow \tilde{e} e \rightarrow 
\tilde{G} e^+e^-$, as predicted by SUSYGEN, were taken into 
account in the calculation of the exclusion regions. These 
branching ratios affected in a significant way only the limit at
low neutralino masses under the $m_{\tilde{e}}$~=~75~GeV/c$^2$ 
assumption.

The second scenario was the no-scale
supergravity model where the selectron masses and the neutralino
composition depend on the neutralino mass. The no-scale
supergravity model predicts a
very light gravitino and the cross-section limit and the exclusion
region for this model in the $m_{\tilde{\chi^0_1}}$ versus $m_{\tilde{G}}$ 
mass plane are shown in Figure~\ref{nlz}c.

\begin{figure}[hbt]
\begin{center}
\mbox{\epsfxsize 10cm \epsfbox{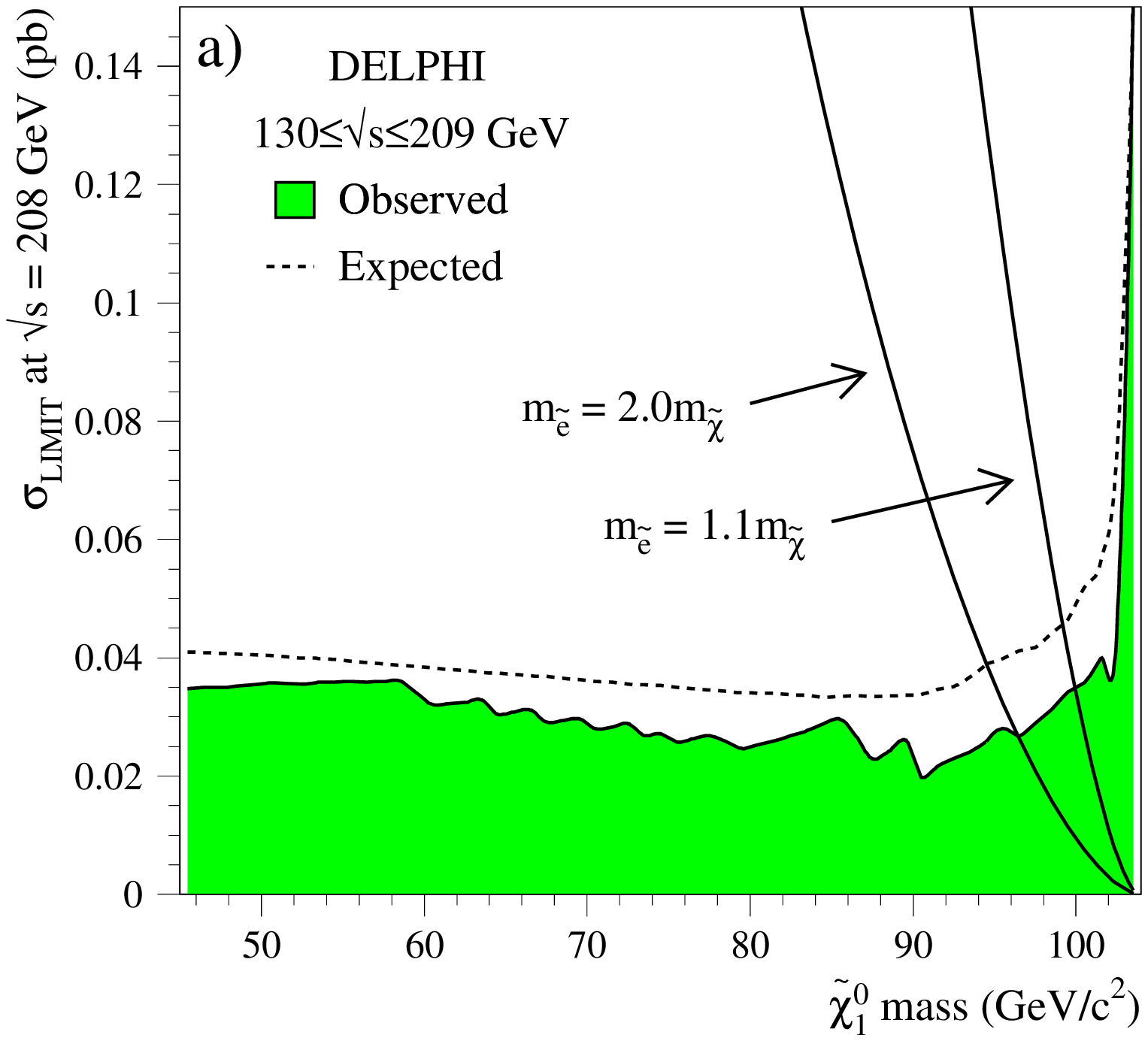}}
\mbox{\epsfxsize 10cm \epsfbox{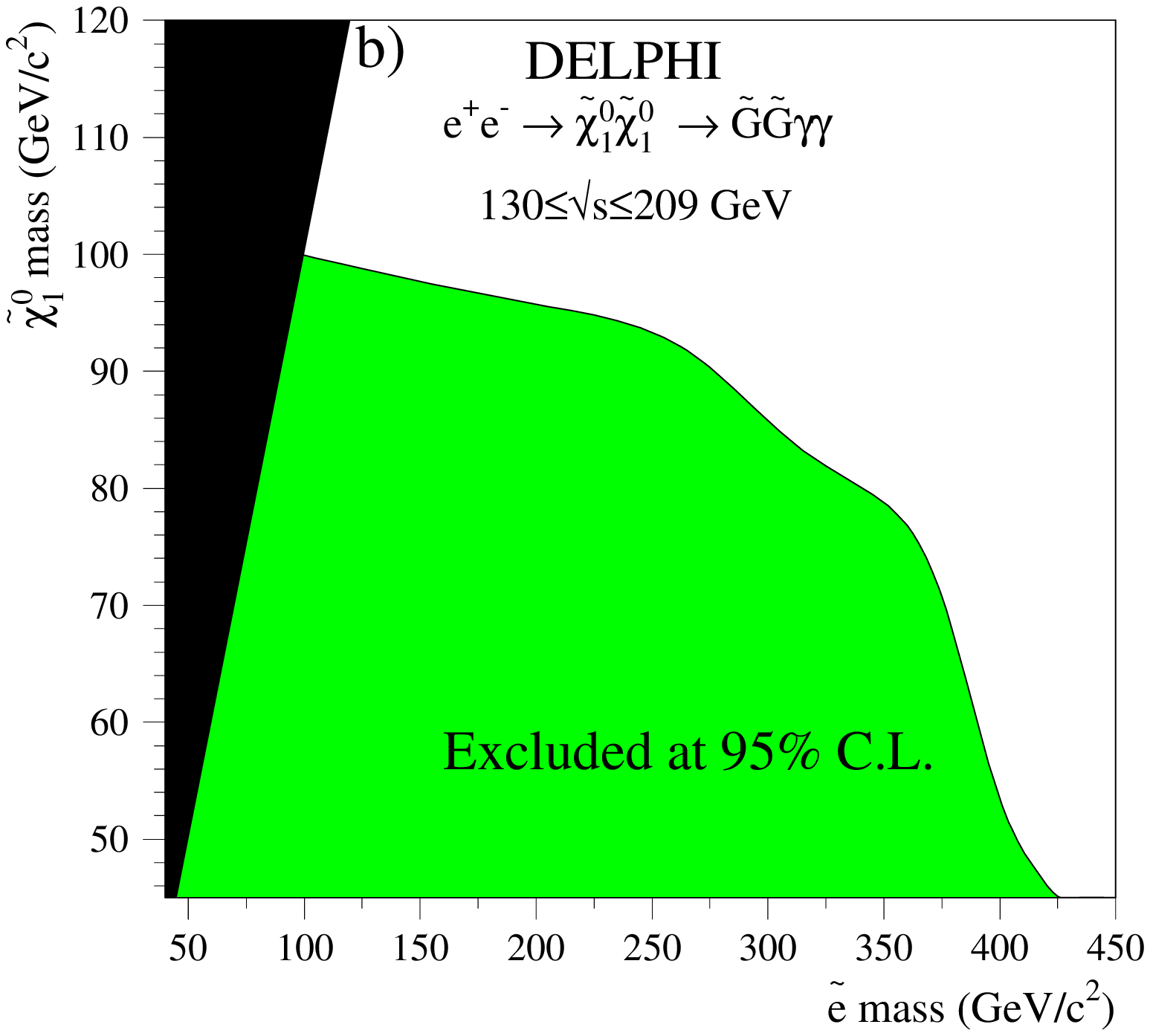}}
\end{center}
\caption[]{a) Upper limit at 95\%~C.L. on the cross-section at $\sqrt{s}=$208~GeV of the process
$\eeGgGg$ as a function of the $\tilde{\chi}^0_1$ mass and the predicted
cross-section for two different assumptions for the selectron mass.
The limit was obtained by combining all data
taken at $\sqrt{s}=$130-209~GeV, assuming the signal cross-section scales as
$\beta/s$ (where $\beta$ is the neutralino velocity).
b) The shaded area shows the
exclusion region in the $m_{\tilde{\chi}}$ versus $m_{\tilde{e}_R}$ plane,
calculated from the DELPHI data at $\sqrt{s}=$130-209~GeV.
} 
\label{chi_yyyy}
\end{figure}

In the search for $\eeGgGg$ at $\sqrt{s}=$190-209~GeV, 17 events were
observed with 13.8 expected from Standard Model sources.  This brings
the total number of events found at $\sqrt{s}=$130-209~GeV to 24 with
20.9 expected (Table~\ref{samples}). 
Figure~\ref{chi_yyyy} shows the
cross-section limit~\cite{LR} calculated from these events as a
function of the $\tilde{\chi}^0_1$ mass (assuming a branching ratio of
100\% for $\tilde{\chi}^0_1\rightarrow\tilde{G}\gamma$) and the
exclusion region in the $m_{\tilde{\chi}}$ versus $m_{\tilde{e}_R}$
plane. A total systematic uncertainty
of $\pm$5\% was assumed for the signal efficiency, which included the
uncertainties on the signal simulation ($\pm$4\%) and on the trigger
efficiency ($\pm$3\%). This was taken into account in deriving the 
cross-section limit but it had a negligible effect on the result.   
 
The dependence of the signal cross-section on the selectron
mass in Figure~\ref{chi_yyyy} is due to the possibility of t-channel 
selectron exchange in the
production mechanism.  A lower
limit of 96~GeV/c$^2$ (100~GeV/c$^2$)~at~95\% C.L. for the $\tilde{\chi}^0_1$
mass can be deduced under the assumption that the neutralino is a pure
bino and with the hypotheses
$m_{\tilde{e}_R}=m_{\tilde{e}_L}=2m_{\tilde{\chi}}$
($m_{\tilde{e}_R}=m_{\tilde{e}_L}=1.1m_{\tilde{\chi}}$).

If the gravitino mass is larger than 200-300~eV/c$^2$, the
$\tilde{\chi}^0_1$ can have such a long lifetime that it will decay
far from the production point yet within the detector.  The signature
for this case is photons that do not point to the interaction region.
If the decay length is long, the probability to detect both photons is
small and it is more efficient to search for the signal events in a 
non-pointing single-photon sample. In this analysis it was therefore 
required that the photon had a
shower axis reconstructed in the HPC which gave a beam crossing point
at least 40~cm away from the interaction point.  Ten events were found
at 190-209~GeV with 7.0 expected, bringing the total at all energies to
16 with 14.6 expected from Standard Model sources (Table~\ref{samples}).

\begin{figure}[hbt]
\centerline{\epsfig {file=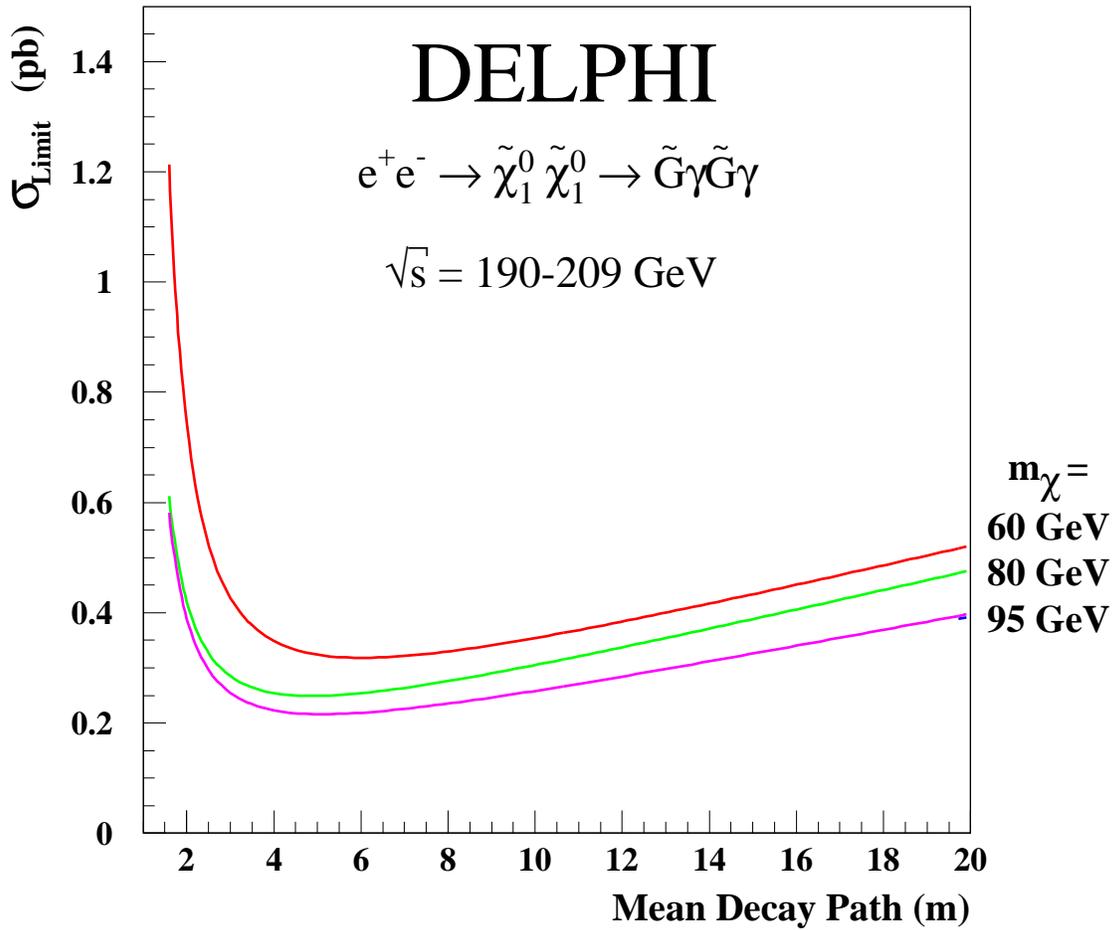,width=15cm}}
\caption[]{
Upper limit at 95\%~C.L. on the cross-section
of the process $\eeGgGg$
as a function of the $\tilde{\chi}^0_1$ mean decay path for different
hypotheses for the neutralino mass. The data collected at $\sqrt{s}=$190-209~GeV
were used for this plot.
} 
\label{chi_zz} 
\end{figure}

Figure~\ref{chi_zz} shows the cross-section limit as a function of the
mean decay path of the neutralino from the non-pointing single-photon
events. The limit was calculated under the assumptions that the neutralino
is a pure Bino, produced dominantly via right-handed selectron t-channel
exchange and that it decays isotropically to $\tilde{G}\gamma$. The ratio
$m_{\tilde{e}}/m_{\tilde{\chi}}$ was furthermore assumed to be between
1.1 and 2.0 and no significant difference in signal efficiency was observed
within this window.

\subsubsection{Limits on neutralino production if the $\tilde{\chi}^0_1$ is the LSP}

In other SUSY models~\cite{SUGRA} the $\tilde{\chi}^0_1$ is the LSP
and $\tilde{\chi}^0_2$ is the NLSP.  The $\eeXgXg$ process has an
experimental signature which is the same as for $\eeGgGg$ but with
somewhat different kinematics due to the masses of the
$\tilde{\chi}^0_1$ and $\tilde{\chi}^0_2$.  The previous DELPHI
analyses at lower energies~\cite{delphi_189,delphi_gg} have now been
repeated with the 190-209~GeV data sample. 14 events remain after
all cuts, with 15.2 expected from the Standard Model background
(Table~\ref{samples}).  The total number of events found at
$\sqrt{s}=$130-209~GeV was 26 with 23.8 expected. The expected photon
distributions from $\tilde{\chi}^0_1$ were again predicted by
SUSYGEN, giving a flat distribution in $cos(\theta)$ and in
energy, with minimum and maximum energies which depended on the masses
of the particles involved in the reaction (and on $\sqrt{s}$).

\begin{figure}[hbt]
\begin{center}
\mbox{\epsfxsize 12cm \epsfbox{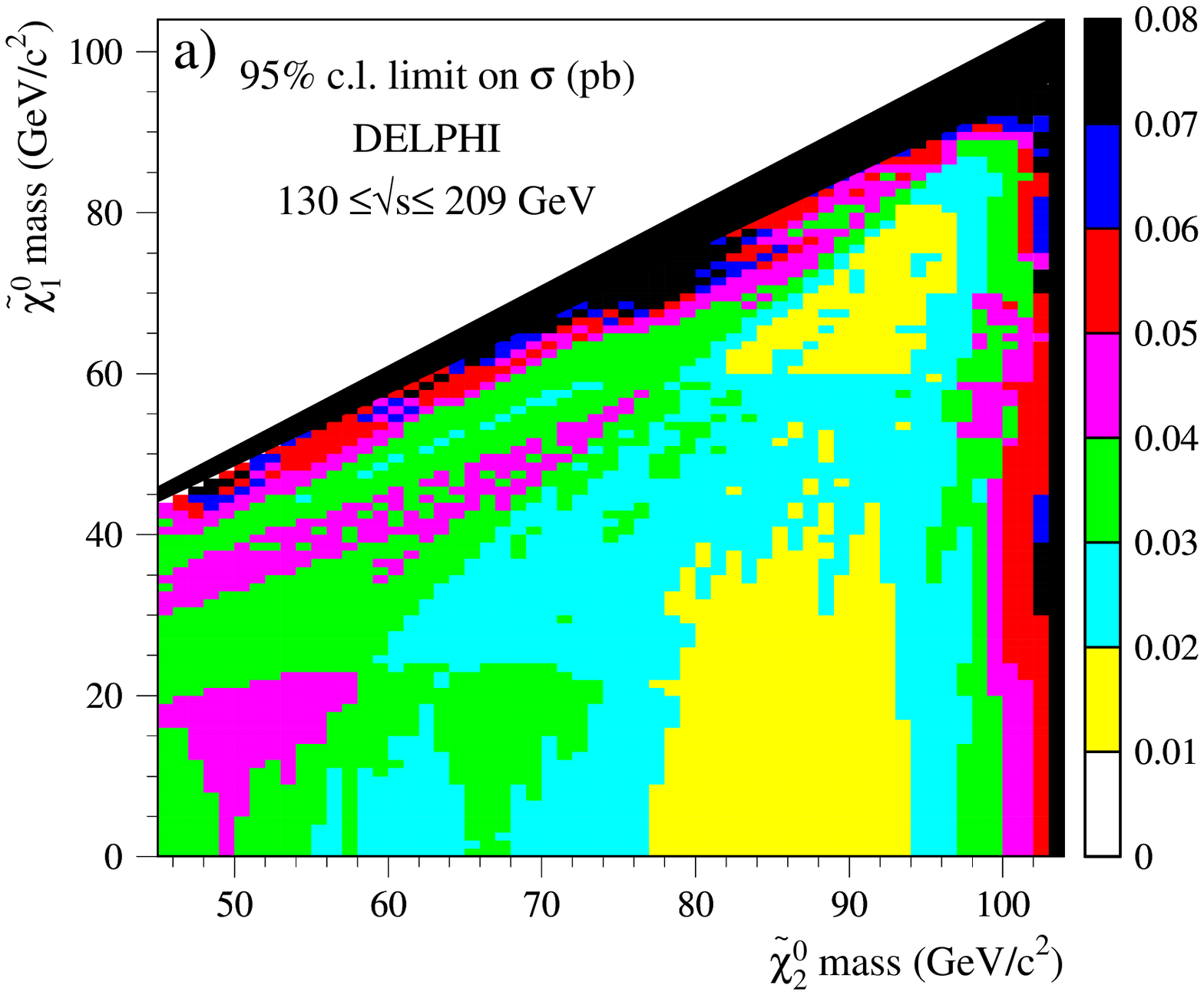}}
\mbox{\epsfxsize 12cm \epsfbox{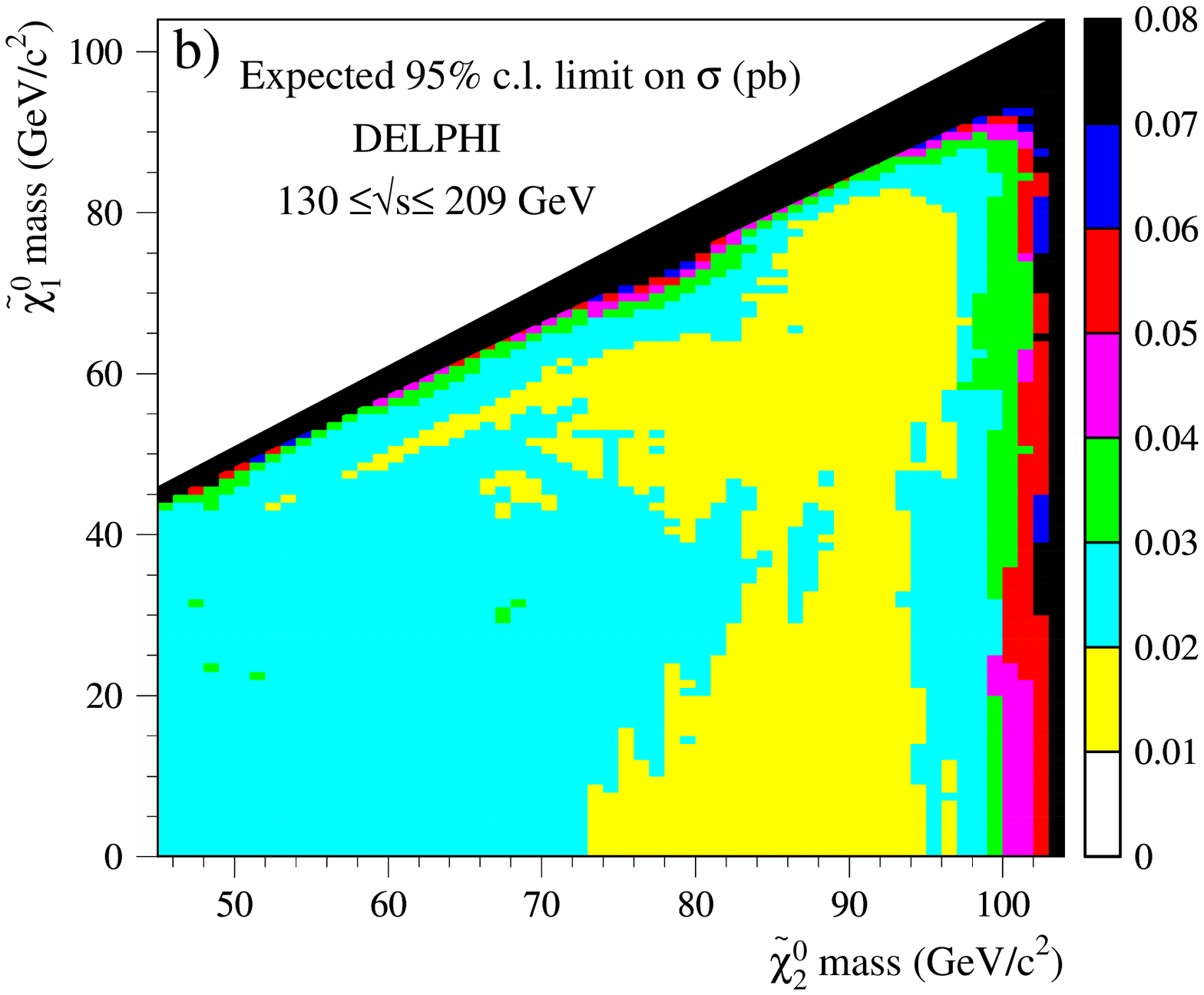}}
\end{center}
\caption[]{a) The observed upper limit at 95\%~C.L. on the cross-section 
at $\sqrt{s}=$208~GeV of the process $\eeXgXg$ as a function of the $\tilde{\chi}^0_1$ 
and the $\tilde{\chi}^0_2$ mass.
The different shaded areas correspond to limits in pb as indicated by the shading scale on the 
right hand side. The limit was obtained by combining the data taken at $\sqrt{s}=$130-209~GeV, 
assuming the signal cross-section to scale as $\beta/s$.
b) The expected upper limit at 95\%~C.L. on the cross-section at $\sqrt{s}=$208~GeV of the process
$\eeXgXg$.
} 
\label{chi_yy} 
\end{figure}

Figure~\ref{chi_yy}
shows the observed and expected cross-section limits calculated from
the events collected at all energies as a function of the
$\tilde{\chi}^0_1$ and $\tilde{\chi}^0_2$ masses, assuming a branching
ratio of 100\% for
$\tilde{\chi}^0_2\rightarrow\tilde{\chi}^0_1\gamma$.

\section{Conclusions}

The DELPHI experiment has analysed all single- and acoplanar
multi-photon events collected during 1995-2000 at a centre-of-mass
energy between 130-209~GeV.

The measured single- and multi-photon cross-sections are in agreement
with the expectations from the Standard Model process $e^+e^-
\rightarrow \nu \bar{\nu}\gamma(\gamma)$ and the number of light 
neutrino generations was measured to be $N_{\nu}=2.84\pm0.10(stat)\pm0.14(syst)$.

The absence of an excess of events has been used to set model-independent
limits on the production of new neutral states, a light
gravitino and neutralinos.

New limits on the gravitational scale and on non-standard model
interactions have also been determined.

\input{acknow.tex}

\clearpage

\end{document}

%% file: acknow.tex
\subsection*{Acknowledgements}
\vskip 3 mm
 We are greatly indebted to our technical 
collaborators, to the members of the CERN-SL Division for the excellent 
performance of the LEP collider, and to the funding agencies for their
support in building and operating the DELPHI detector.\\
We acknowledge in particular the support of \\
Austrian Federal Ministry of Education, Science and Culture,
GZ 616.364/2-III/2a/98, \\
FNRS--FWO, Flanders Institute to encourage scientific and technological 
research in the industry (IWT), Belgium,  \\
FINEP, CNPq, CAPES, FUJB and FAPERJ, Brazil, \\
Czech Ministry of Industry and Trade, GA CR 202/99/1362,\\
Commission of the European Communities (DG XII), \\
Direction des Sciences de la Mati$\grave{\mbox{\rm e}}$re, CEA, France, \\
Bundesministerium f$\ddot{\mbox{\rm u}}$r Bildung, Wissenschaft, Forschung 
und Technologie, Germany,\\
General Secretariat for Research and Technology, Greece, \\
National Science Foundation (NWO) and Foundation for Research on Matter (FOM),
The Netherlands, \\
Norwegian Research Council,  \\
State Committee for Scientific Research, Poland, SPUB-M/CERN/PO3/DZ296/2000,
SPUB-M/CERN/PO3/DZ297/2000, 2P03B 104 19 and 2P03B 69 23(2002-2004)\\
FCT - Funda\c{c}\~ao para a Ci\^encia e Tecnologia, Portugal, \\
Vedecka grantova agentura MS SR, Slovakia, Nr. 95/5195/134, \\
Ministry of Science and Technology of the Republic of Slovenia, \\
CICYT, Spain, AEN99-0950 and AEN99-0761,  \\
The Swedish Research Council,      \\
Particle Physics and Astronomy Research Council, UK, \\
Department of Energy, USA, DE-FG02-01ER41155. \\
EEC RTN contract HPRN-CT-00292-2002. \\
